\documentclass[superscriptaddress,twocolumn]{revtex4-2}
\pdfoutput=1
\usepackage[utf8]{inputenc}

\usepackage{bbold}
\usepackage{amsmath, amssymb, graphicx, bm}
\usepackage{physics}
\usepackage{soul}
\usepackage[newfloat,frozencache,cachedir=minted-cache]{minted}
\usepackage{hyperref}
\usepackage[capitalise]{cleveref}
\usepackage[labelformat=simple]{subcaption}

\setcitestyle{numbers,sort&compress}

\usepackage{array,booktabs}
\newcolumntype{M}[1]{>{\centering\arraybackslash}m{#1}} 
\usepackage{makecell}

\usepackage{bbding} 
\usepackage{rotating}
\usepackage{tikz}

\newcommand{\minn}{m_\text{io}}
\newcommand{\mout}{m_\text{io}}
\newcommand{\mio}{m_\text{io}}
\newcommand{\mfb}{m_\text{mem}}
\newcommand{\mfbenc}{m_\text{mem,enc}}
\newcommand{\mfbdec}{m_\text{mem,dec}}

\newcommand{\Ob}{\mathbf{O}}
\newcommand{\Oenc}{\mathbf{O}_\text{enc}}
\newcommand{\Odec}{\mathbf{O}_\text{dec}}
\newcommand{\Sb}{\mathbf{S}}
\newcommand{\Ienc}{I_{\text{enc}}}

\newcommand{\phienc}{\phi_{\text{enc}}}
\newcommand{\phidec}{\phi_{\text{dec}}}
\newcommand{\thetaenc}{\theta_{\text{enc}}}
\newcommand{\thetadec}{\theta_{\text{dec}}}

\newcommand{\D}{D}

\usetikzlibrary{shapes.geometric}

\newcommand{\redstar}{%
  \tikz{\node[star, star points=5, star point height=0.5em, draw=black, fill=red, scale=0.3, rotate=36]{};}%
}

\begin{document}

\title{A Quantum Optical Recurrent Neural Network\\for Online Processing of Quantum Times Series}

\author{Robbe De Prins}
\email{robbe.deprins@ugent.be}
\affiliation{Photonics Research Group, Ghent University – imec, Technologiepark-Zwijnaarde 126, 9052 Gent, Belgium.}

\author{Guy Van der Sande}
\email{guy.van.der.sande@vub.be}
\affiliation{Applied Physics Research Group, Vrije Universiteit Brussel, Pleinlaan 2, 1050 Brussels, Belgium.}

\author{Peter Bienstman}
\email{peter.bienstman@ugent.be}
\affiliation{Photonics Research Group, Ghent University – imec, Technologiepark-Zwijnaarde 126, 9052 Gent, Belgium.}

\begin{abstract}
    Over the last decade, researchers have studied the synergy between quantum computing (QC) and classical machine learning (ML) algorithms. However, measurements in QC often disturb or destroy quantum states, requiring multiple repetitions of data processing to estimate observable values. 
    In particular, this prevents online (i.e., real-time, single-shot) processing of \textit{temporal} data as measurements are commonly performed during intermediate stages.
    Recently, it was proposed to sidestep this issue by focusing on tasks with quantum output, thereby removing the need for detectors.
    Inspired by reservoir computers, a model was proposed where only a subset of the internal parameters are optimized while keeping the others fixed at random values \cite{nokkala2021onlineprocessing}.
    Here, we also process quantum time series, but we do so using a quantum optical recurrent neural network (QORNN) of which \textit{all} internal interactions can be trained.
    As expected, this approach yields higher performance, as long as training the QORNN is feasible. 
    We further show that our model can enhance the transmission rate of quantum channels that experience certain memory effects. 
    Moreover, it can counteract similar memory effects if they are unwanted, a task that could previously only be solved when redundantly encoded input signals were available.
    Finally, we run a small-scale version of this last task on the photonic processor Borealis \cite{Borealis}, demonstrating that our QORNN can be constructed using currently existing hardware.
\end{abstract}

\maketitle

\section{Introduction}

In the pursuit of improved data processing, there is an increasing emphasis on combining machine learning (ML) techniques with quantum computing (QC). Building on the established belief that quantum systems can outperform classical ways of computing \cite{harrow2017quantum}, quantum machine learning (QML) \cite{cerezo2022challenges} investigates how to design and deploy software and hardware that harnesses the complexity of such systems. Inverse to this `QC for ML' approach, also research is being carried out to improve QC algorithms and hardware using ML techniques.

In classical machine learning, algorithms such as recurrent neural networks (RNNs) \cite{salehinejad2017recent,medsker1999recurrent}, transformers \cite{transformer,RNNvsTransformer}, long short-term memory (LSTM) networks \cite{yu2019review}, and reservoir computing (RC) \cite{Guy2017advances} have led to state-of-the-art performances in natural language processing, computer vision, and audio processing. This makes them good sources of inspiration for new QML models.

However, 
the common use of projective measurements in quantum computing leads to the requirement of processing the same input data multiple times to estimate the expectation values of detected observables. This is a real bottleneck for \textit{temporal} tasks, as such measurements are often carried out at intermediate processing stages, leading to back-actions on the state of the quantum system. On the one hand, this leads to laborious operating procedures and large overheads \cite{mujal2021opportunities}. On the other hand, it prevents one from performing online time series processing (i.e.~constantly generating output signals in real-time, based on a continuous stream of input signals).

Recently, an approach was introduced that proposes to sidestep this detection issue by performing online processing of quantum states and thereby removing the need for detectors \cite{nokkala2021onlineprocessing}. 
The model was inspired by the concept of RC \cite{Guy2017advances}, where random dynamical systems, also called reservoirs, are made to process temporal input data.
RC research has demonstrated that training only a simple output layer to process the reservoir's output signals can achieve state-of-the-art performance in various computational tasks while significantly reducing training costs.
Building on this idea, Ref.~\cite{nokkala2021onlineprocessing} tackled several computational tasks using a random network of harmonic oscillators and training only the interactions between that network and some input-carrying oscillators.

Here, we introduce a quantum optical recurrent neural network (QORNN) in which all interactions are trainable within the Gaussian state formalism \cite{olivares2012quantum}.
We first compare our model with the findings of Ref.~\cite{nokkala2021onlineprocessing} by conducting classical simulations of two computational tasks: the short-term quantum memory (STQM) task and the entangler task. We will provide detailed definitions of these tasks in the following sections. They respectively serve as benchmarks to assess the QORNN's linear memory capabilities and its ability to entangle different states in a time series. 
The results will demonstrate that the relaxation of the RC strategy to a QORNN indeed increases the performance, as long as training is tractable.

We further demonstrate the capabilities of our model by applying it to a quantum communication task. Specifically, we show that the QORNN can enhance the capacity of a quantum memory channel. In such a channel, subsequent signal states are correlated through interactions with the channel's environment. Our network achieves this enhancement by generating an entangled quantum information carrier.
Indeed, it is known that the asymptotic transmission rate of memory channels can be higher than the maximal rate achieved by separable (unentangled) channel uses. It is said that the capacity of such channels is `superadditive'. 
For a bosonic memory channel with additive Gauss-Markov noise \cite{schafer2009capacity}, it was previously shown that the generation of such entangled carriers can be performed sequentially (i.e.~without creating all channel entries all at once) while achieving near-optimal enhancement of the capacity \cite{schafer2012GMPS}. 
Our model achieves the same result, while having a simpler encoding scheme and being more versatile, as it can adapt to hardware imperfections.

Moreover, we show that a QORNN can also compensate for unwanted memory effects in quantum channels (the so-called quantum channel equalization or QCE task). Existing work on this task required the availability of redundantly encoded input signals \cite{nokkala2021onlineprocessing}. This undermines the practicality of the method. Moreover, such a redundant encoding is impossible without full prior knowledge of the input states (e.g.~when they result from a previous quantum experiment that is not exactly repeatable) because quantum states cannot be cloned. Here, we show that the increased flexibility of the QORNN allows us to lift the restriction of redundant encoding. Additionally, we find that the QORNN's performance can be improved by allowing the reconstruction of the channel's input to be performed with some delay.

Finally, we run a small-scale version of the QCE task on the recently introduced photonic processor Borealis \cite{Borealis}. Although the results are restricted by the limited tunability of Borealis' phase modulators, we can still demonstrate that our model can be constructed using currently existing hardware. 

The rest of this paper is structured as follows. In \cref{sec: model}, we introduce our QORNN model. In \cref{sec:benchmark tasks}, we benchmark our model with the STQM task (\ref{sec:STQM}) and the entangler task (\ref{sec:entangler}). In \cref{sec:superadditivity}, we show how the QORNN can lead to superadditivity in a bosonic memory channel.
In \cref{sec:QCE:redundancy}, we show that the QORNN can tackle the QCE task without requiring its input signals to be encoded with redundancy. Finally, the results of the experimental demonstration of the QCE task are given in \cref{sec:QCE:borealis}.

\section{Model} 
\label{sec: model}
\begin{figure}[!htb]
	\centering
	\includegraphics[width=0.5\textwidth]{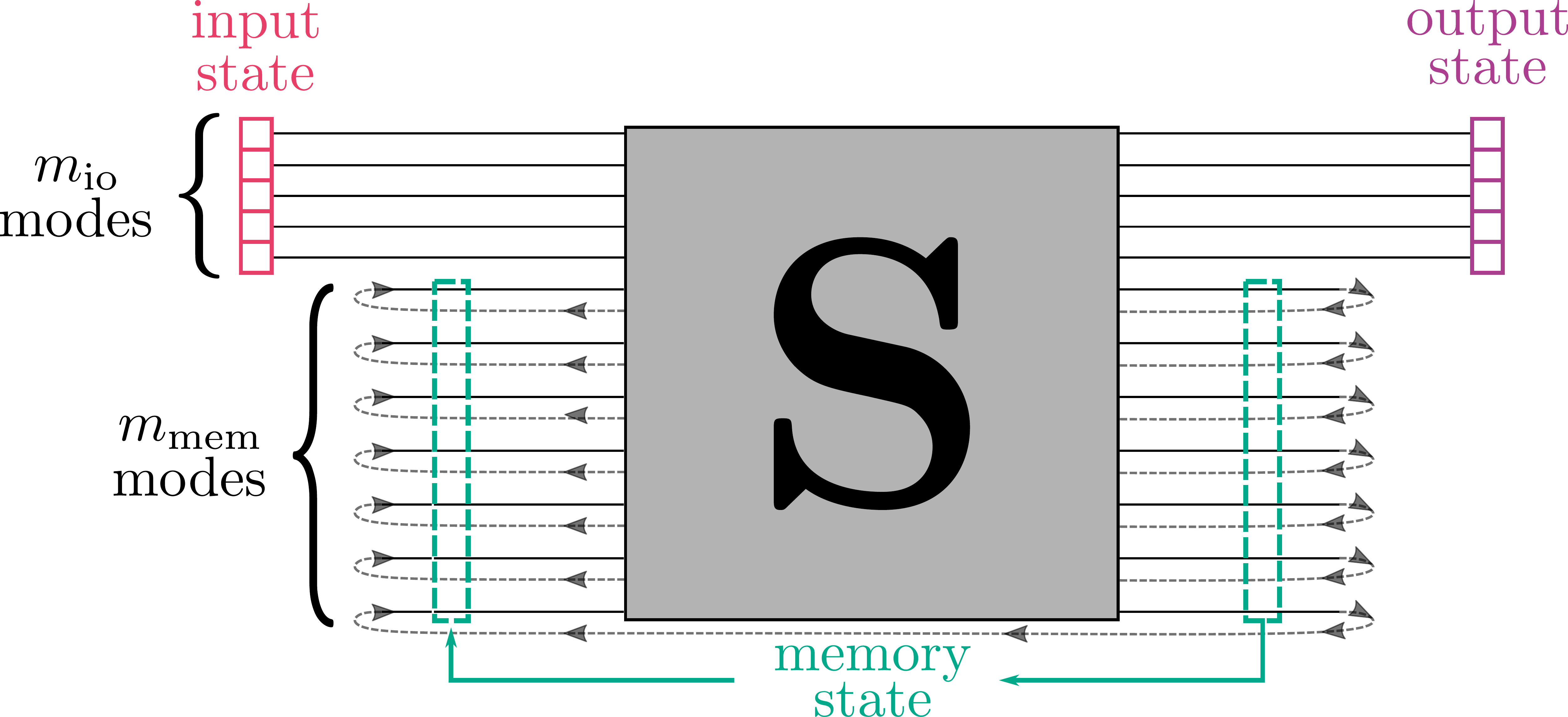}
	\caption{QORNN model. Quantum states are repeatedly sent in the upper $\minn$ modes. These input modes are combined with $\mfb$ memory modes and sent through a symplectic network $\Sb$ (i.e.~a network of beam splitters, phase shifters, and optical squeezers). Afterwards, the state on the upper $\mout$ modes is collected as output, while the remaining $\mfb$ modes are looped back to the left side of $\Sb$. 
    }
	\label{fig: QRNN model}
\end{figure}

Our QORNN model is presented in \cref{fig: QRNN model}. 
It incorporates an $m$-mode circuit $\Sb$ that generally consists of beam splitters, phase shifters, and optical squeezers. Such a circuit can be described by a symplectic matrix. Hence, we will further refer to it as a symplectic circuit.
The $\mio$ upper modes at the left (right) side of $\Sb$ are the input (output) modes of the QORNN. The remaining modes of $\Sb$ are connected from left to right using $\mfb = m - \minn$ delay lines. The delay lines are equally long and we fill further denote them as `memory modes'.
To perform a temporal task, we send a time series of quantum states (e.g., obtained from encoding classical information or a previous quantum experiment) to the input modes of the QORNN. The temporal spacing between the states is chosen equal to the propagation time of $\Sb$ and the delay lines, such that we can describe the QORNN operation in discrete time.
Because of the memory modes, output states depend on multiple past input states, which grants the QORNN some memory capacity. By training the circuit $\Sb$ (essentially training the parameters of its constituent gates), the QORNN can learn to process temporal data. 

In further sections, we sometimes restrict $\Sb$ to be \textit{orthogonal} symplectic. Such a circuit comprises only beam splitters and phase shifters, with optical squeezers being excluded. When applicable, we will denote it as $\mathbf{O}$.

\section{Benchmark  tasks}
\label{sec:benchmark tasks}
\subsection{Short-term quantum memory task}
\label{sec:STQM}

\begin{figure}[!htb]
	\centering
	\includegraphics[width=\linewidth]{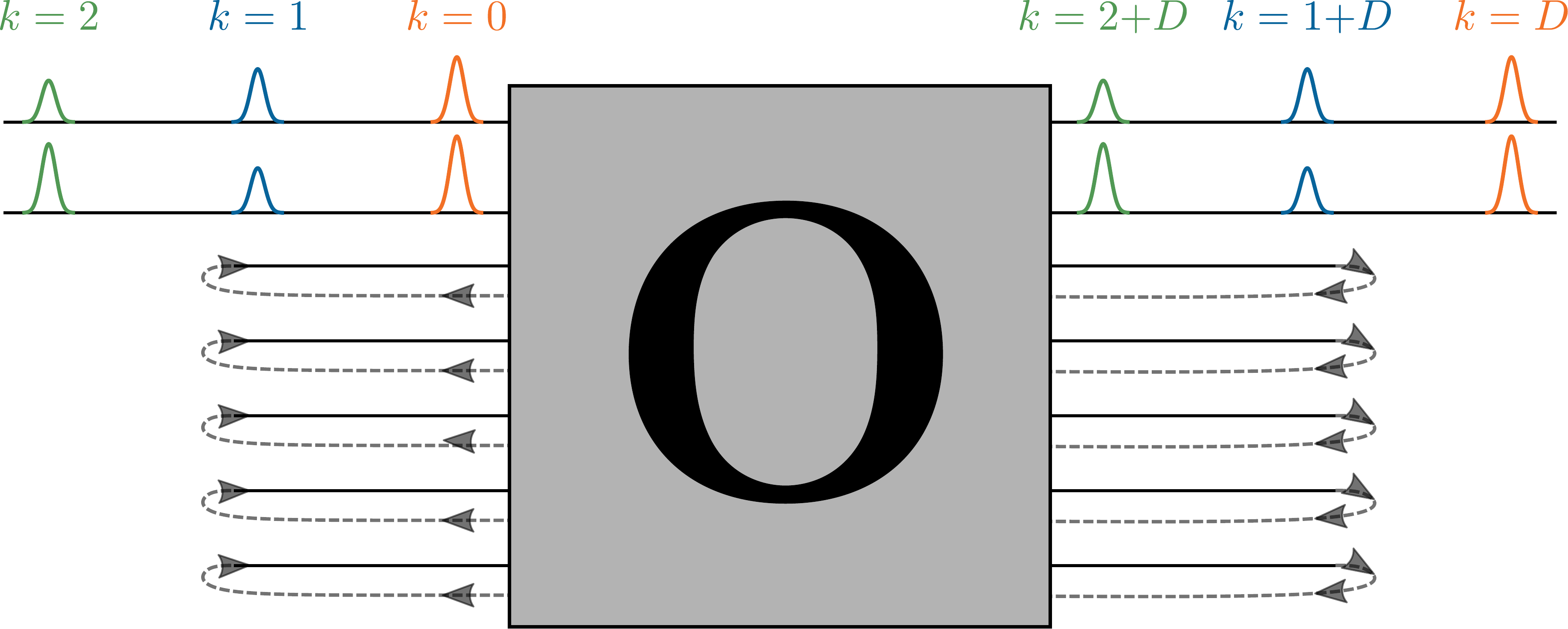}
	\caption{Setup for the STQM task with $\mout=2$. The QORNN consists of an orthogonal symplectic network $\Ob$. Pulses of different colors represent a time series of quantum states. A state that is sent into the QORNN at iteration $k$ should appear at the output at iteration $k+\D$.}
	\label{fig:STQM setup}
\end{figure}

The goal of the short-term quantum memory (STQM) task is to recall states that were previously fed to the QORNN after a specific number of iterations, denoted by $\D$. This task is visualized in \cref{fig:STQM setup} for the case where $\mout=2$ and the QORNN consists of an orthogonal circuit. 
Note that if we were to use a general symplectic network instead of an orthogonal one, optical squeezers could be added and optimized, such the results would be at least equally good. However, we will show that we can reach improved performance without including optical squeezers in the QORNN, which is beneficial for an experimental setup.

\begin{figure}[!htb]
\centering
\begin{subfigure}{1\linewidth}
\caption{}
  \includegraphics[width=\linewidth]{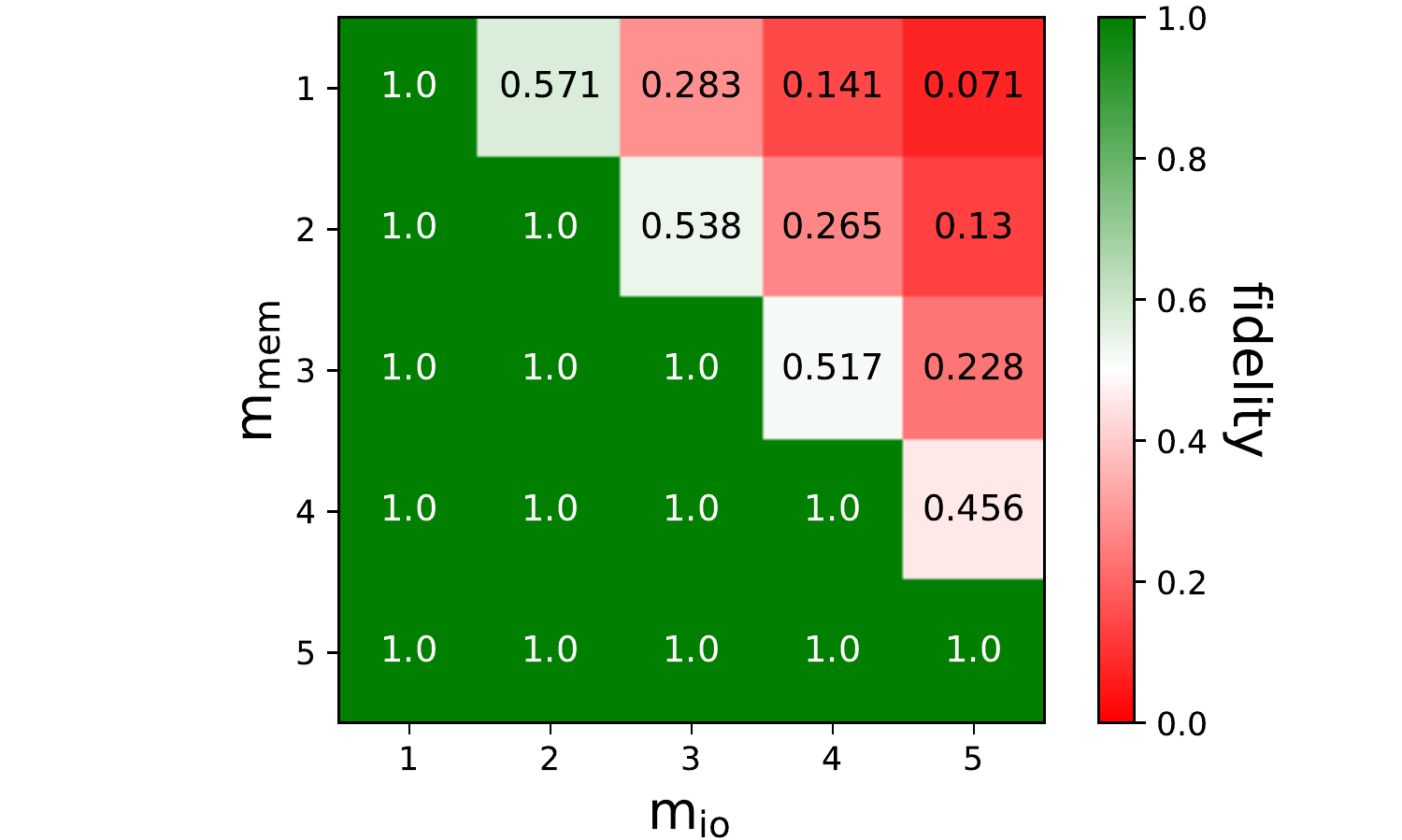}
  \label{fig:STQM this work}
\end{subfigure}
\begin{subfigure}{1\linewidth}
\caption{}
  \includegraphics[width=\linewidth]{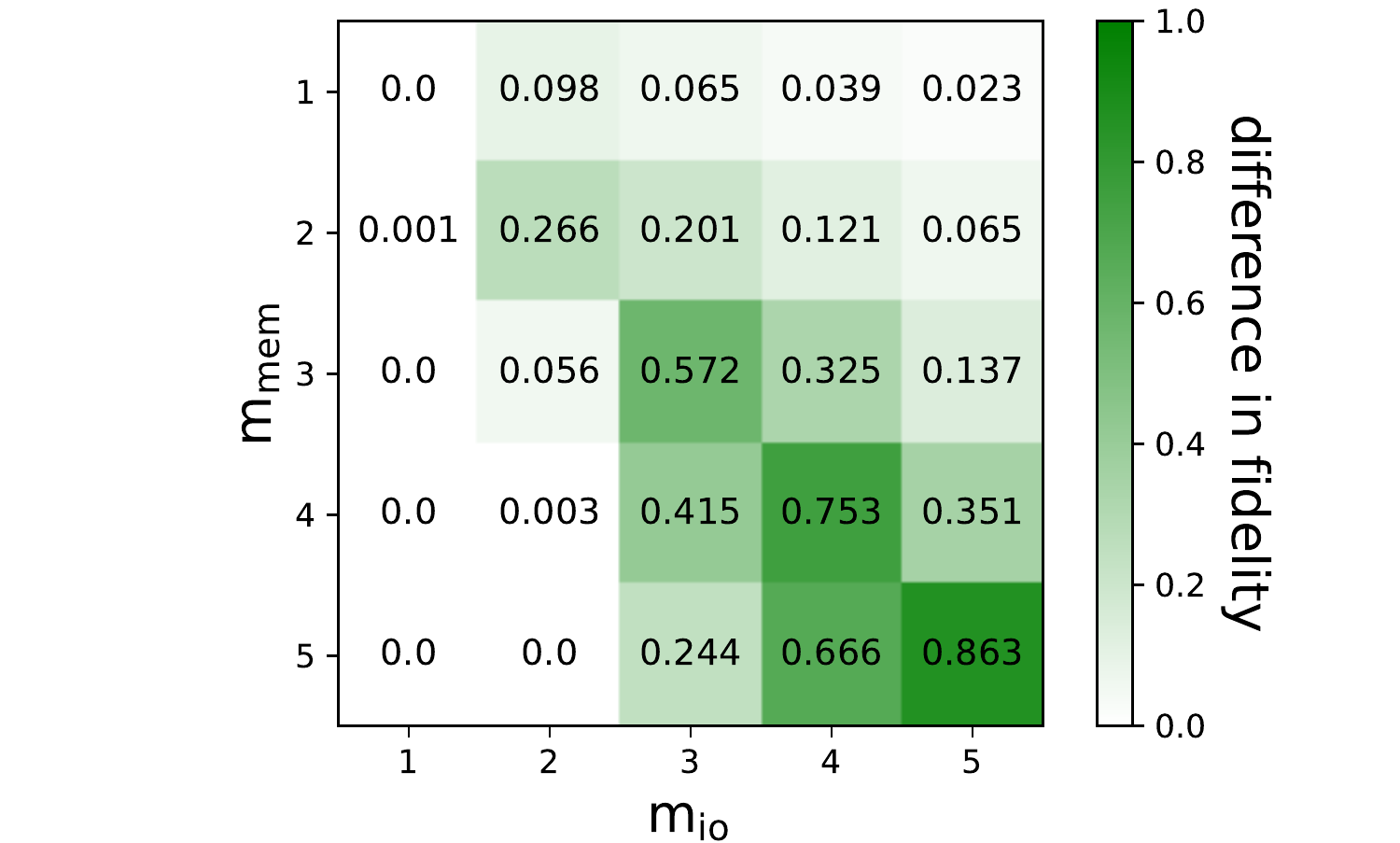}
  \label{fig:STQM comparison}
\end{subfigure}
\caption{STQM performance for $\D=1$ and for different values of $\mout$ and $\mfb$.
 Fig.~(a) shows the average fidelity between a desired output state and a state resulting from the QORNN. In Fig.~(b), the corresponding results achieved in Ref.~\cite{nokkala2021onlineprocessing} are subtracted from our results.
}
\label{fig: STQM sweep}
\end{figure}

We focus our attention on the case where $\D=1$. 
The input states are chosen randomly from a set of squeezed thermal states (more details in \hyperref[sec:methods:STQM]{Section A of Methods}).
\cref{fig:STQM this work} shows the average fidelity between an input state at iteration $k$ and an output state at iteration $k+\D$, as a function of $\mfb$ and $\mout$. We see that the QORNN perfectly solves the STQM task if $\mout \leq \mfb$. 
This is easy to understand as $\Ob$ can be trained to perform several SWAP operations (i.e.~operations that interchange the states on two different modes). More specifically, the QORNN can learn to swap every input mode with a different memory mode, such that the input state is memorized for a single iteration before being swapped back to the corresponding output mode.
For $\mout > \mfb$, such a SWAP-based circuit is not possible, leading to less than optimal behavior of the QORNN.

In \cref{fig:STQM comparison}, the fidelity values obtained by the RC-inspired model of  Ref.~\cite{nokkala2021onlineprocessing} are subtracted from our results.
Across all values of $\mfb$ and $\mout$, we observe that the QORNN scores equally well or better.
Although Ref.~\cite{nokkala2021onlineprocessing} also achieves a fidelity of 1 for certain combinations of $\mout$ and $\mfb$, the set of these combinations is smaller than for the QORNN.  
Moreover, it is important to note that the RC-inspired design limits the number of trainable parameters, making a SWAP-based solution impossible in general.
As a result, prior to training, it is more challenging to guarantee optimal performance of the RC-inspired model, while this is not the case for the QORNN.

\subsection{Entangler task}
\label{sec:entangler}

\begin{figure}[!htb]
	\centering
	\includegraphics[width=\linewidth]{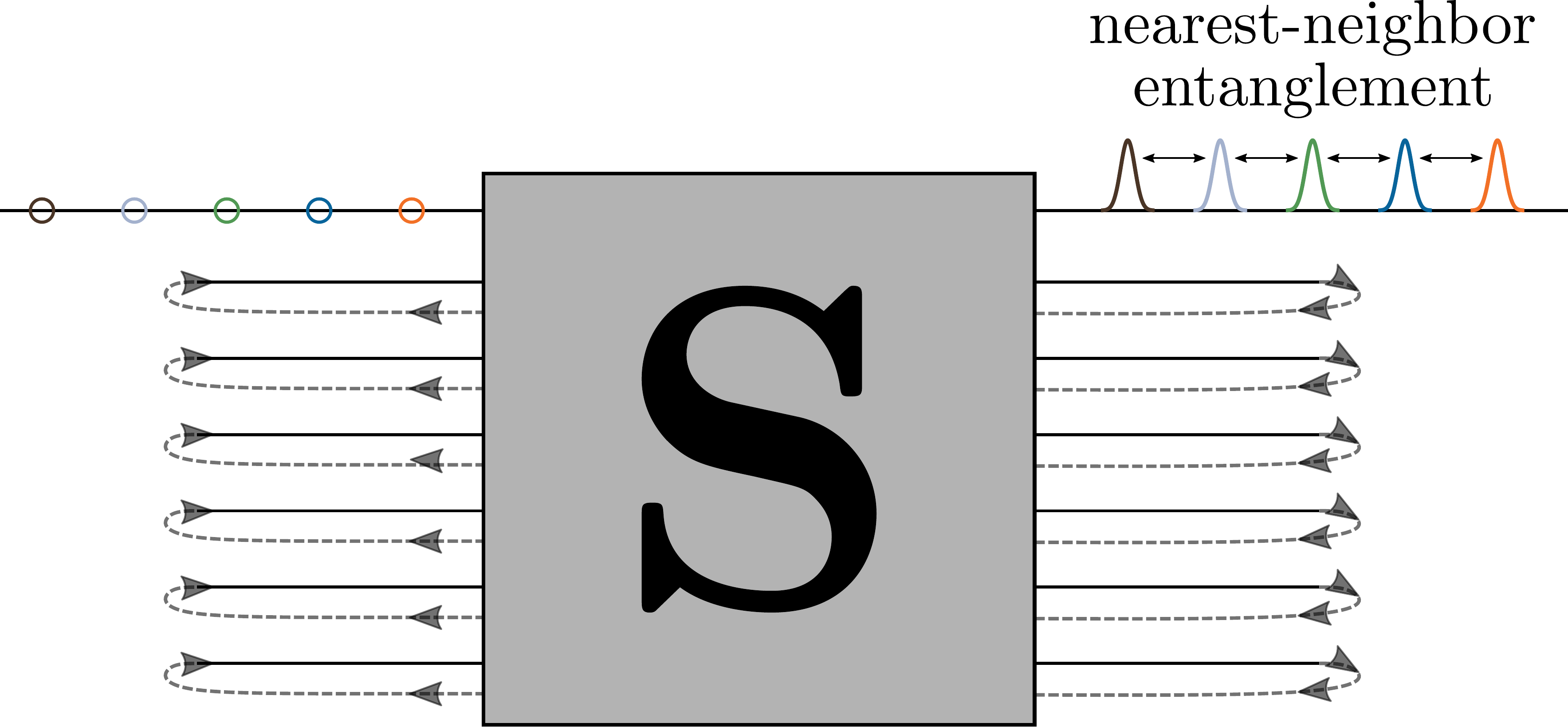}
	\caption{Setup for the entangler task for $\mout=1$ and spacing $S=1$. Circles of different colors represent an input time series of vacuum states. Pulses of different colors are entangled output states.}
	\label{fig:ENT setup}
\end{figure}

The objective of the entangler task is to entangle different states of a time series that were initially uncorrelated. The performance of this task is evaluated based on the average logarithmic negativity between output states at iterations $k$ and $k+S$.
Negativity \cite{olivares2012quantum} is an entanglement measure for which higher values indicate greater levels of entanglement between the states.
Note that if we consider output states with spacing $S=1$, then we aim to entangle nearest-neighbor states.
This last task is visualized in \cref{fig:ENT setup} for the case where $\mout=1$. We choose vacuum states as input and hence the circuit $\Sb$ should \textit{not} be orthogonal as we want to generate states with nonzero average photon numbers.

\begin{figure}[!htb]
\centering
\begin{subfigure}{1\linewidth}
\caption{}
  \includegraphics[width=\linewidth]{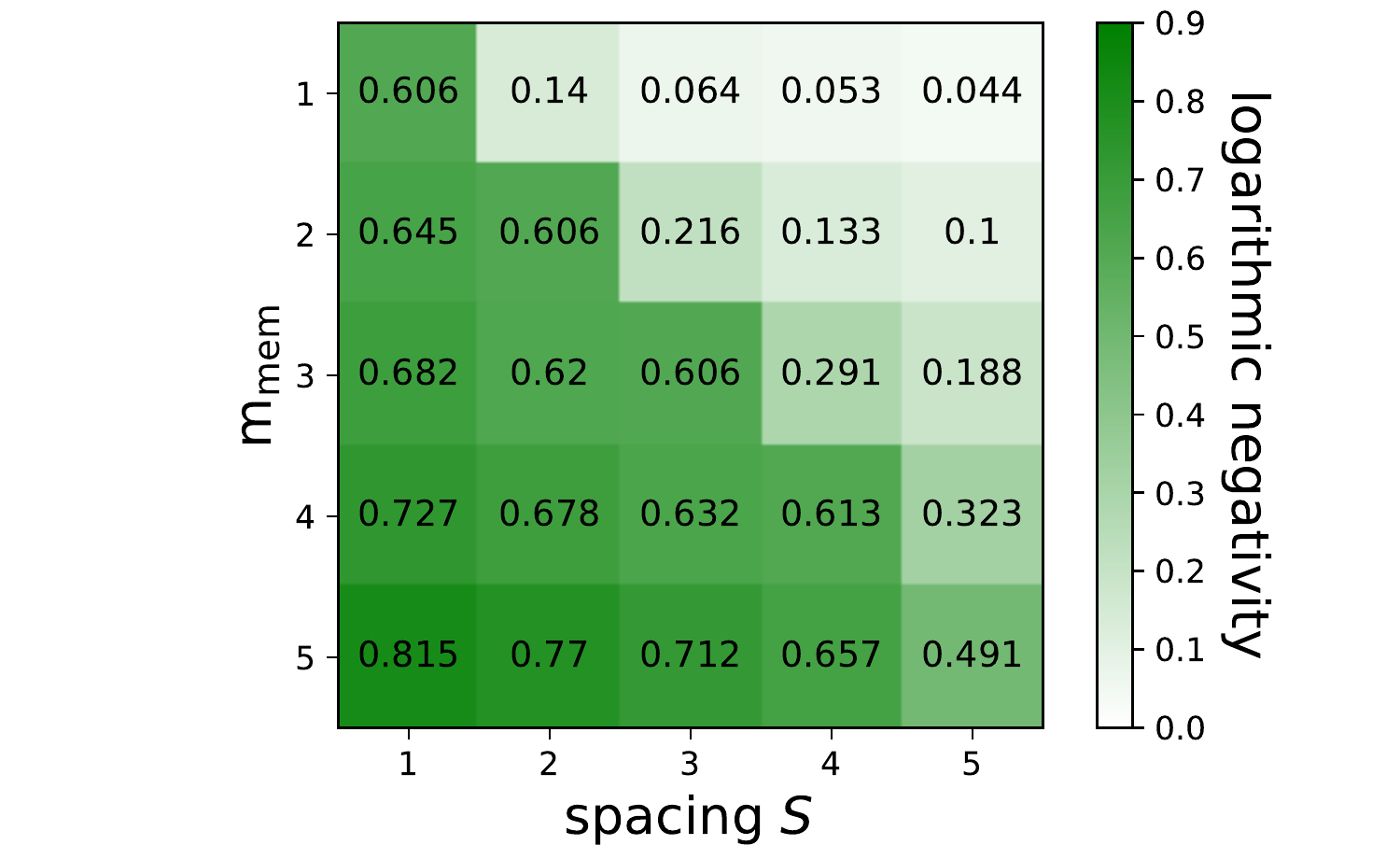}
  \label{fig:ENT_results:this work}
\end{subfigure}
\begin{subfigure}{1\linewidth}
\caption{}
  \includegraphics[width=\linewidth]{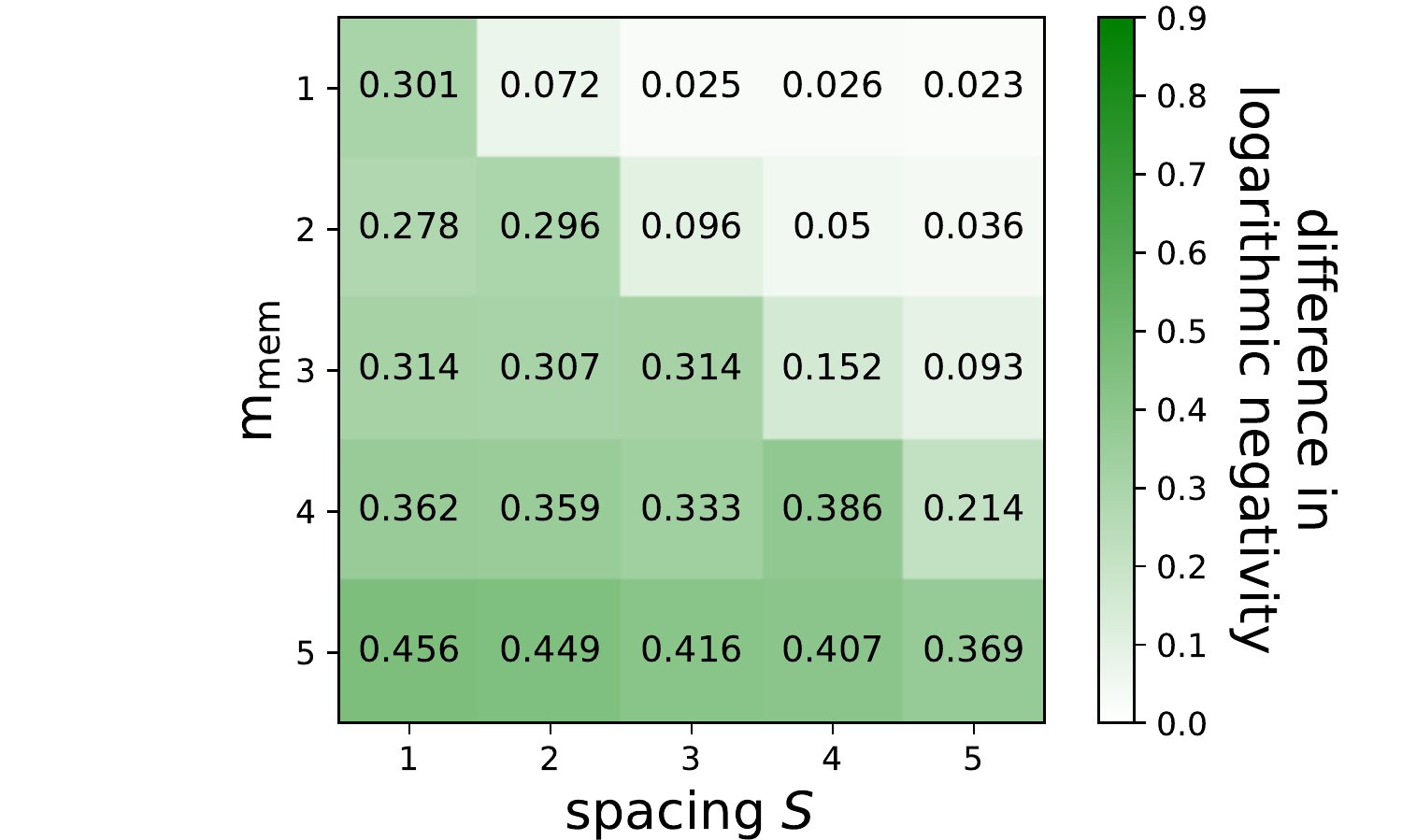}
  \label{fig:ENT_results:comparison}
\end{subfigure}
\caption{Entangler task performance for $\minn=1$ and for different values of $S$ and $\mfb$. Fig.~(a) shows the logarithmic negativity resulting from the QORNN. In Fig.~(b), the corresponding results achieved in Ref.~\cite{nokkala2021onlineprocessing} are subtracted from our results.\cite{nokkala2021onlineprocessing}.}
\label{fig:ENT_results}
\end{figure}

For $\mout=1$, \cref{fig:ENT_results:this work} displays the average logarithmic negativity obtained by the QORNN for various values of $\mfb$ and $S$. 
For a given spacing, the performance increases with $\mfb$. This can be attributed to the fact that a higher value of $\mfb$ leads to a bigger circuit $\Sb$, such that more entangling operations can be applied.
It can also be seen that the performance roughly stays the same along the diagonal ($S=\mfb$) and along lines parallel to the diagonal. This can be explained by the findings of Section \ref{sec:STQM}, which indicate that increasing $\mfb$ can effectively address the increased linear memory requirements of the task that arise from increasing $S$.
Finally, our comparison with the RC-inspired model proposed in Ref.~\cite{nokkala2021onlineprocessing}, as shown in \cref{fig:ENT_results:comparison}, indicates that the QORNN performs better, owing to its larger number of trainable parameters. 

\section{Quantum communication  tasks}
\label{sec:comm tasks}

\subsection{Superadditivity}
\label{sec:superadditivity}

In this section, we show that the QORNN can enhance the transmission rate of a quantum channel that exhibits memory effects.
When a state is transmitted through such a `memory channel', it interacts with the channel's environment. As subsequent input states also interact with the environment, correlations arise between different channel uses. 
Contrary to memoryless channels, it is known that the transmission rate of memory channels can be enlarged by providing them with input states that are entangled over subsequent channel uses \cite{caruso2014quantum}, a phenomenon that is better known as `superadditivity'. 
Here, we aim to create such entangled input states using our QORNN. 

Note the similarity with the definition of the entangler task. Now however, the goal is not to create maximal entanglement between the different states, but rather a specific type of entanglement that depends on the memory effects of the channel and that will increase the transmission rate.

\begin{figure*}[!htb]
	\centering
	\includegraphics[width=\linewidth]{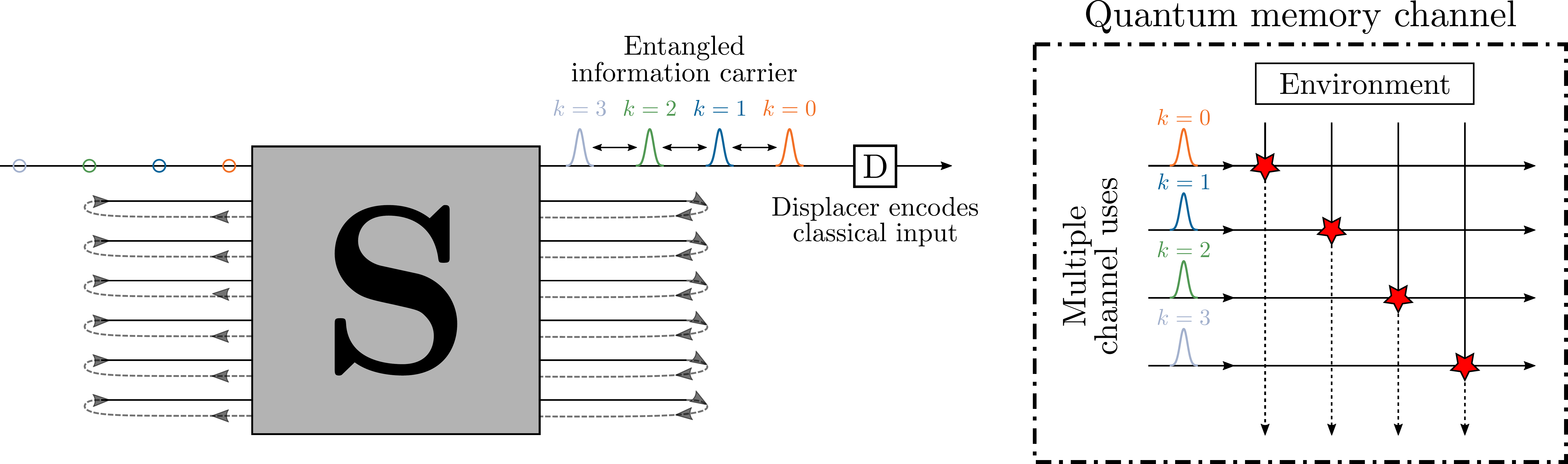}
	\caption{Setup for the superadditivity task. A QORNN (with $\minn=1$) transforms vacuum states into a quantum information carrier that is entangled across different time bins. A displacer (D) modulates this carrier to encode classical input information. The resulting signal is sent to a bosonic memory channel \cite{schafer2009capacity}. A number of $K$ consecutive uses of the channel are modeled as a single parallel K-mode channel.
 The channel's environment introduces noise (\protect\redstar) that leads to correlations between the successive channel uses. As a result of the entangled carrier, the transmission rate of the channel can be enhanced.}
	\label{fig:superadditivity_setup}
\end{figure*}

The setup for the `superadditivity task' is shown in \cref{fig:superadditivity_setup}.
A QORNN with $\minn = 1$ transforms vacuum states into an entangled quantum time series. 
Information is encoded by displacing each individual state of the series over a continuous distance in phase space. These distances are provided by a classical complex-valued information stream. Their probabilities follow a Gaussian distribution with zero mean and covariance matrix $\bm{\gamma}_{\mathrm{mod}}$. 

The resulting time series is then sent through a memory channel. 
A number of $K$ consecutive uses of the channel are modeled as a single parallel K-mode channel.
The memory effects we consider here are modeled by correlated noise emerging from a Gauss-Markov process \cite{schafer2009capacity}. The environment has the following classical noise covariance matrix $\bm{\gamma}_{\mathrm{env}}$:
\begin{gather}
    \bm{\gamma}_{\mathrm{env}} = \left(\begin{array}{cc}\mathbf{M}(\phi) & 0 \\ 0 & \mathbf{M}(-\phi)\end{array}\right), \label{eq: env noise cov matrix}\\[0.3em]
    M_{i j}(\phi) = N \phi^{|i-j|} ~. \label{eq: env noise cov matrix (M)}
\end{gather}
Here, $\phi \in [0, 1)$ denotes the strength of the nearest-neighbor correlations and $N \in \mathbb{R}$ is the variance of the noise. In \cref{eq: env noise cov matrix}, $\mathbf{M}(\phi)$  correlates the $q$ quadratures, while $\mathbf{M}(-\phi)$ anti-correlates the $p$ quadratures.

The transmission rate of the channel is calculated from the von Neumann entropy of the states that pass through the channel (i.e.~from the Holevo information). Here we adopt the approach and the parameter values outlined in Ref.~\cite{schafer2009capacity}.

Note that the average photon number that is transmitted per channel use ($\bar{n}$) has a contribution from both the QORNN (i.e.~from its squeezers) and from the displacer. Given a value for $\bar{n}$, the transmission rate is maximized by training both the circuit $\Sb$ and $\bm{\gamma}_{\mathrm{mod}}$ under the energy constraint imposed by $\bar{n}$. Nonzero squeezing values are obtained, leading to an information carrier.
This highlights the counter-intuitive quantum nature of the superadditivity phenomenon: by spending a part of the available energy on the carrier generation rather than on classical modulation, one can reach higher transmission rates, something that has no classical analog. 

\begin{figure}[!htb]
	\centering
	\includegraphics[width=0.9\linewidth]{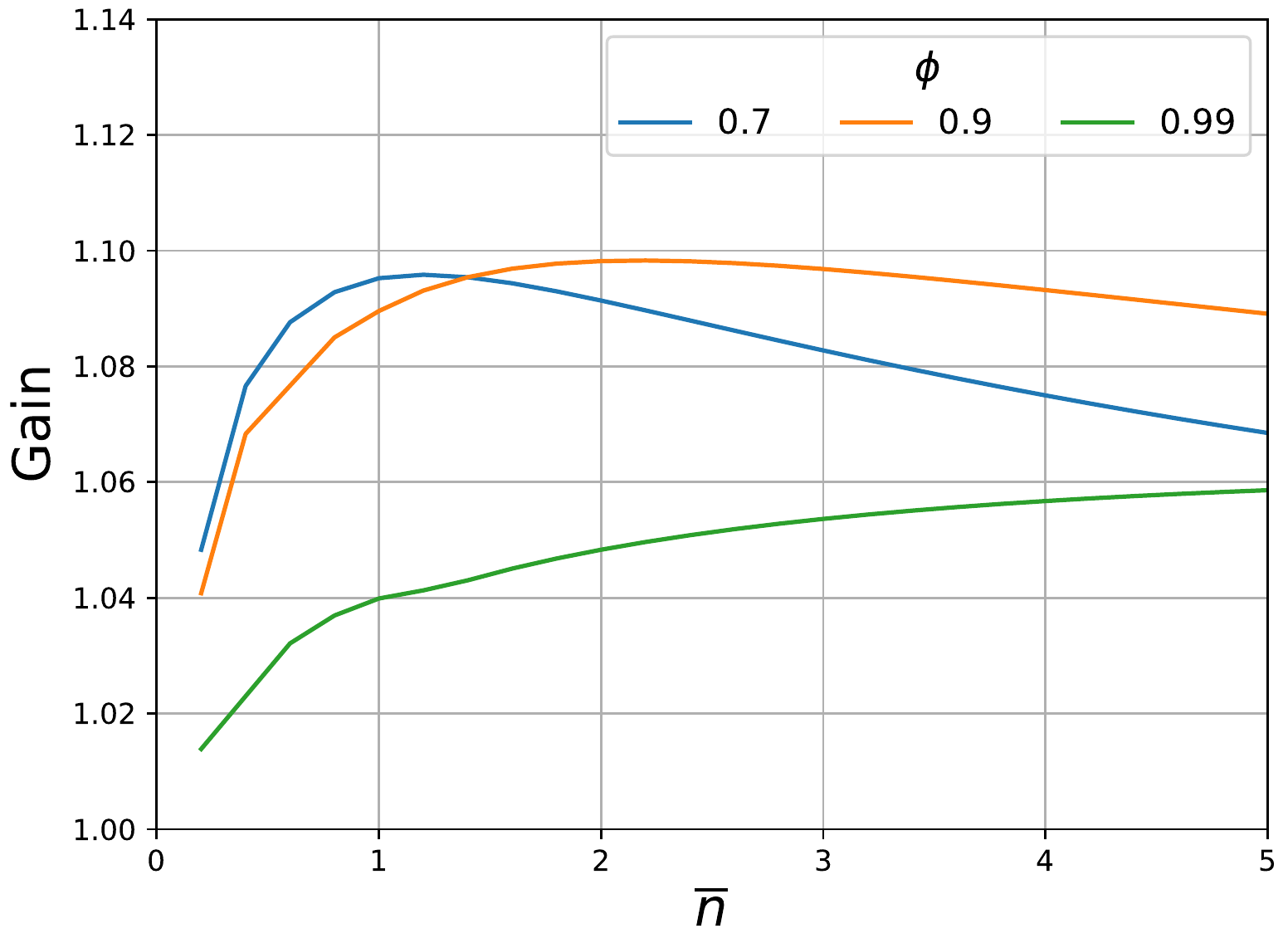}
	\caption{Performance of the superadditivity task for 30 channel uses. The gain in transmission rate is plotted as a function of the average photon number per use of the channel ($\bar{n}$) and for different values of the noise correlation parameter ($\phi$). 
    Additional parameters are chosen as follows: $\mout=\mfb=1$, $N=\bar{n}/3$.}
	\label{fig:superadditivity_results}
\end{figure}

We now define a quality measure for the superadditivity task. The gain $G$ is the ratio of the achieved transmission rate to the optimal transmission rate for separable (i.e.~unentangled) input states. 
For 30 channel uses, \cref{fig:superadditivity_results} shows $G$ as a function of the average photon number $\bar{n}$ per use of the channel and for different values of the correlation parameter $\phi$. We take the signal-to-noise ratio $\mathrm{SNR}=\bar{n}/N=3$, where $N$ is defined in \cref{eq: env noise cov matrix (M)}.
We observe that superadditivity is achieved, as the gain is higher than 1 and can reach as high as 1.10. These results agree with the optimal gain values that were analytically derived in prior studies of this memory channel (cfr. Fig.~7 of Ref.~\cite{schafer2011GaussMarkov2_TypeB}). 

While a scheme already exists to generate carriers sequentially (i.e., generating carriers without creating all channel entries simultaneously) \cite{schafer2012GMPS}, our model also provides a simpler and more versatile alternative. Unlike the existing scheme, our model eliminates the need for Bell measurements, while achieving the same near-optimal gains. Additionally, our model is able to adapt to hardware imperfections, as they can be taken into account during training.

\subsection{Quantum channel equalization}
\label{sec:QCE:redundancy}
In this section, we use the QORNN as a model for a quantum memory channel. This time, we assume its memory effects to be unwanted (unlike \cref{sec:superadditivity}) and compensate for them by sending the channel's output through a second QORNN instance.

\begin{figure*}[!htb]
	\centering
	\includegraphics[width=0.9\linewidth]{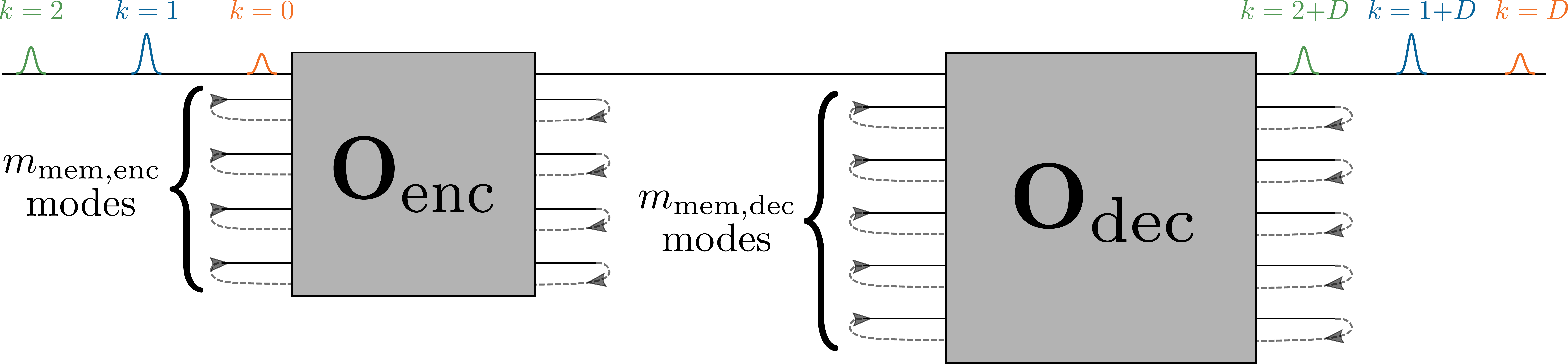}
	\caption{Setup for the QCE task when $\mout=1$ and for a delay $\D$. Pulses of different colors represent a time series of quantum states. The encoder and decoder respectively consist of orthogonal symplectic networks $\Ob_\text{enc}$ and $\Ob_\text{dec}$. $\Ob_\text{enc}$ is initialized randomly and kept fixed. $\Ob_\text{dec}$ is trained such that an input state that is sent into the encoder at iteration $k$ appears at the output of the decoder at iteration $k+\D$.}
	\label{fig:QCE_setup}
\end{figure*} 

\cref{fig:QCE_setup} shows the setup for the quantum channel equalization (QCE) task in more detail. An `encoder' QORNN acts as a model for a memory channel. Because such channels normally do not increase the average photon number of transmitted states, we restrict the encoder's symplectic circuit to be orthogonal and denote it as $\Oenc$. This circuit is initialized randomly and will not be trained later. A second `decoder' QORNN is trained to invert the transformation caused by the encoder. 
Similar to the STQM task, we will show that an orthogonal symplectic circuit $\Odec$ is enough to lead to the desired performance, without requiring optical squeezers, which is beneficial for experimental realizations.
We will further denote the number of memory modes of the encoder and decoder as $\mfbenc$ and $\mfbdec$ respectively.
Finally, we introduce a delay of $\D$ iterations between the input and output time series, similar to the definition of the STQM task (see \cref{fig:STQM setup}).

Assume for a moment that the input time series of the encoder only consists of a single state, i.e.~we are looking at an impulse response of the system. By taking $\D>0$, we  allow the decoder to receive multiple output states from the encoder before it has to reconstruct the original input state.
The longer $D$, the more information that was stored in the memory modes of the encoder will have reached the decoder by the time it needs to start the reconstruction process. A similar reasoning applies when the input time series consists of multiple states.
This approach effectively addresses the challenge posed by the no-cloning principle, which prevents the decoder from accessing information stored in the encoder's memory or in the correlations between the encoder's memory and output.

For the RC-inspired model of Ref.~\cite{nokkala2021onlineprocessing}, only the case where $\D=0$ was considered. The no-cloning problem was addressed by redundantly encoding the input signals of the encoder. I.e., multiple copies of the same state were generated based on \textit{classical} input information and subsequently fed to the model through different modes (`spatial multiplexing') or at subsequent iterations (`temporal multiplexing').
Here, we show that taking $\D>0$ allows us to solve the QCE task without such redundancy, ultimately using each input state only once.
This not only simplifies the operation procedure but also enables us to perform the QCE task without prior knowledge of the input states, which is often missing in real-world scenarios such as quantum key distribution.
As these input states cannot be cloned, our approach significantly increases the practical use of the QCE task.
It is worth noting that such an approach where $D>0$ was also attempted for the RC-inspired model \cite{private_comm_Nokkala}, but this was unsuccessful, which we attribute here to its limited number of trainable interactions. 

Additionally, we will show that the QCE performance of the QORNN is influenced by two key factors: the memory capacity of the decoder (as determined by the value of $\mfbdec$), and the response of the encoder to a single input state (observed at the encoder's output modes).

More formally, we measure the impulse response of the encoder by sending in a single squeezed vacuum state (with an average photon number of $\bar{n}_\text{impulse}$) and subsequently tracking the average photon number $h_{\text{enc}}$ in its output modes over time. We denote the impulse response at iteration $k$ by $h_{\text{enc}}^k$.

We now define:
\begin{equation}
    I_{\text{enc}} = \frac{1}{\bar{n}_\text{impulse}} \sum_{k=0}^{\D} h_{\text{enc}}^k
\end{equation}
$I_{\text{enc}}$ is a re-normalized cumulative sum that represents the fraction of $\bar{n}_\text{impulse}$ that leaves the encoder before the decoder has to reconstruct the original input state.

\begin{figure}[!htb]
\centering
\begin{subfigure}{1\linewidth}
\caption{}
  \includegraphics[width=\linewidth]{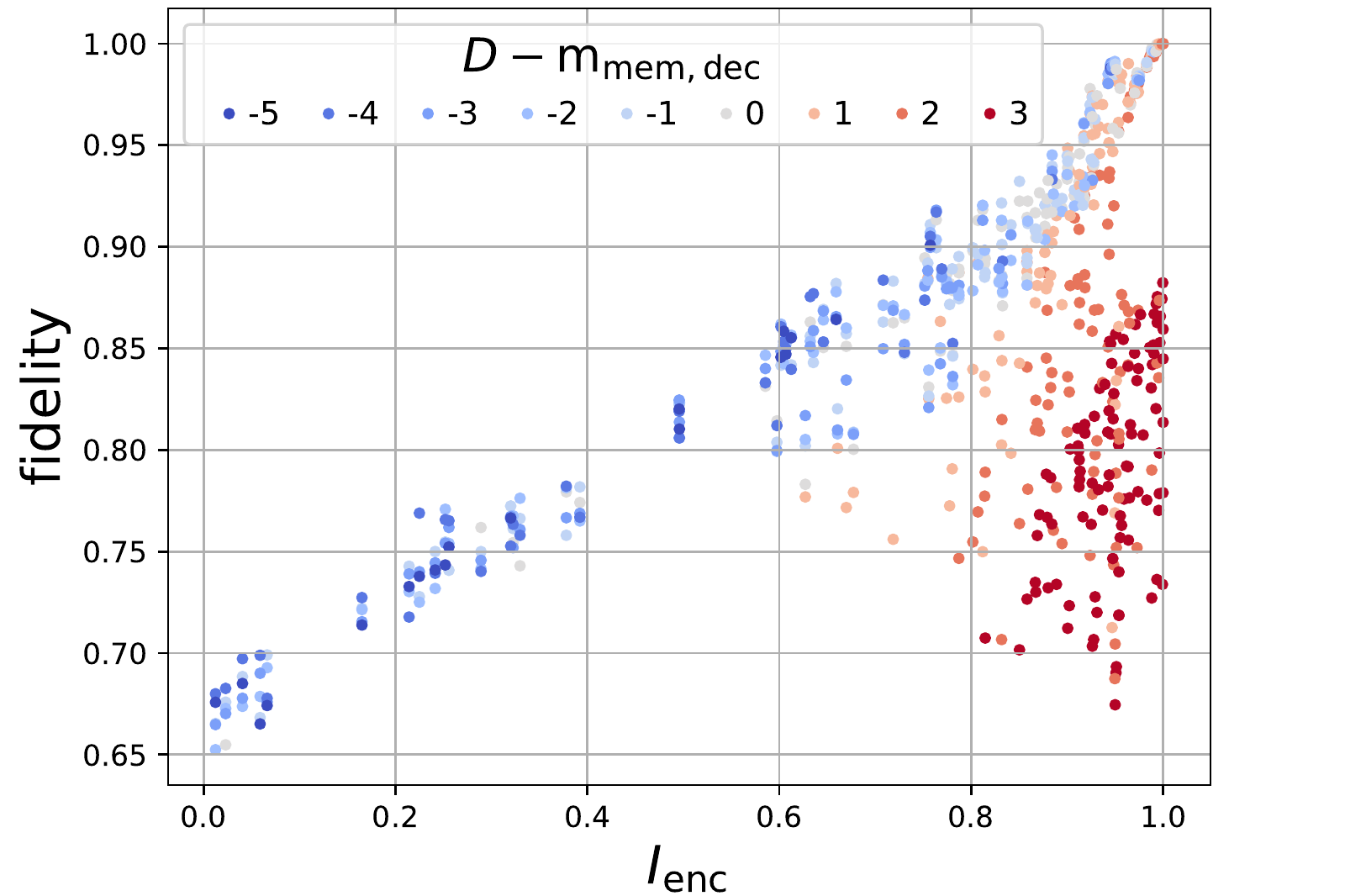}
  \label{fig: QCE fid vs Ienc (multiple tau)}
\end{subfigure}
\begin{subfigure}{1\linewidth}
\caption{}
  \includegraphics[width=\linewidth]{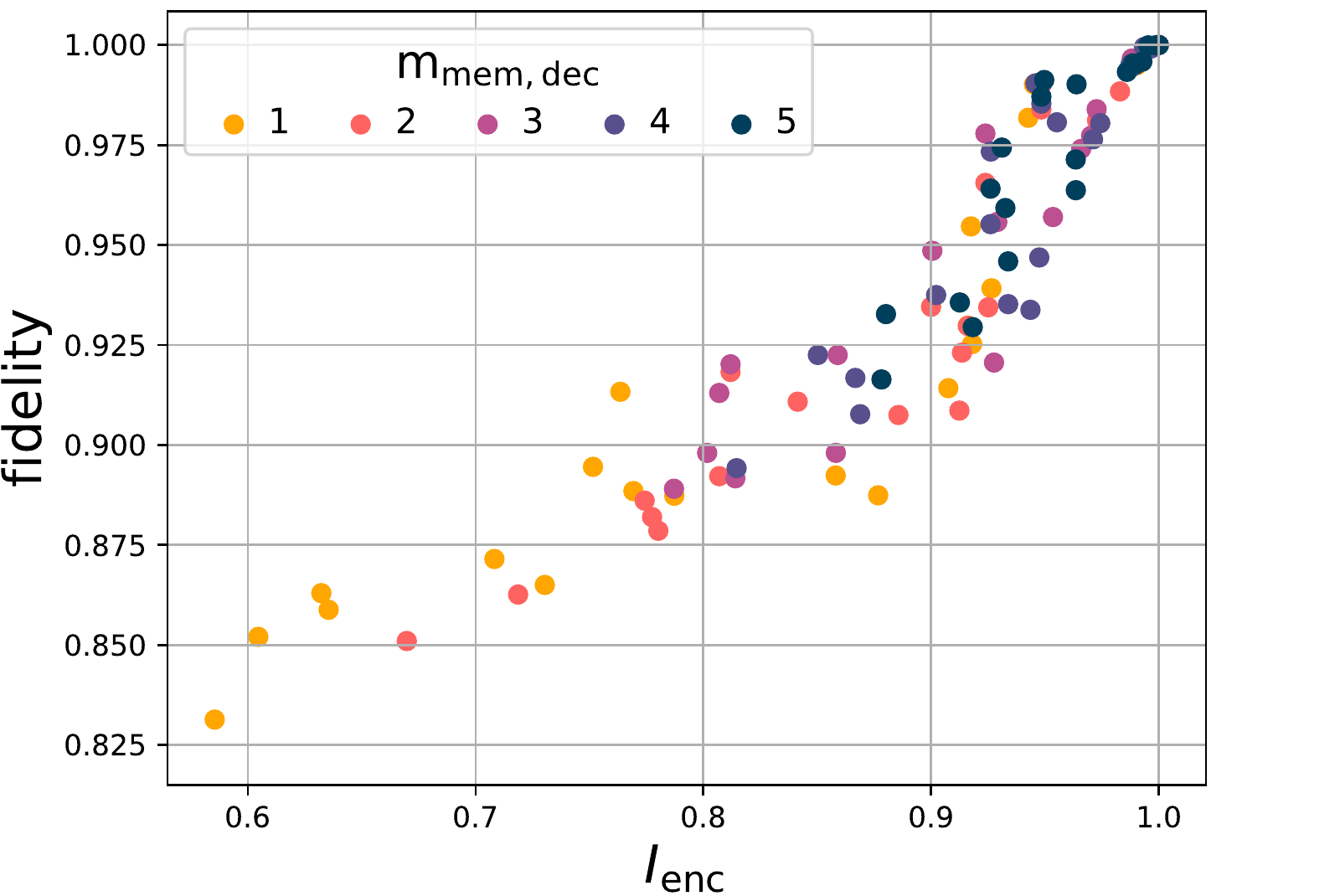}
  \label{fig: QCE fid vs Ienc (best tau)}
\end{subfigure}
\caption{QCE performance for 20 randomly initialized encoders that consist of 2 memory modes.
The results are shown as a function of $I_{\text{enc}}$, i.e.~the fraction of the photon number of a certain input state that reaches the decoder within $\D$ iterations. In Fig.~(a), we consider $\D \in \{0, 1, ...,\mfbdec+\mfbenc+1\}$ and $\mfbdec \in \{1,2,...,5\}$. In Fig.~(b), the optimal value of $\D$ is chosen (given an encoder and $\mfbdec$).
}
\label{fig: QCE fid vs Ienc}
\end{figure}

We now consider 20 randomly initialized encoders with $\mfbenc=2$. 
The input states are randomly sampled from a set of squeezed thermal states (more details in \hyperref[sec:methods:QCE:redundant]{Section D of Methods}).
\cref{fig: QCE fid vs Ienc (multiple tau)} shows the average fidelity between an input state of the encoder at iteration $k$ and an output state of the decoder at iteration $k+\D$ as a function of $\Ienc$ and for different values of $\D - \mfbdec$. We see that if $\D \leq \mfbdec$ (blueish dots), the decoder potentially has enough memory, and the quality of constructing the input states increases as the decoder receives more information from the encoder (i.e.~as $\Ienc$ increases). 
If $\D > \mfbdec$ (reddish dots), we ask the decoder to wait for a longer time before starting to reconstruct the input. This explains why the dots are clustered on the right side of the diagram because more information about the input will be received and $\Ienc$ will be higher. On the other hand, if the delay is too long, it will exceed the memory of the decoder, and the input will start to be forgotten. This explains that the darkest dots with the longest delay have the worst performance. 

Note that $\D$ is a hyperparameter that can be chosen freely. Also note that the optimal choice for the value of $\D$ is not necessarily
$\D = \mfbdec$ (light grey dots) and the actual optimum depends on the exact configuration of the encoder.
\cref{fig: QCE fid vs Ienc (best tau)} shows a subset of the results in \cref{fig: QCE fid vs Ienc (multiple tau)}, where the optimal value of $\D$ is chosen for every encoder initialization and every value of $\mfbdec$. 
We observe that the task can be tackled without redundantly encoded input signals and that the performance increases with both $\mfbdec$ and $\Ienc$. For $\mfbdec=3$, all 20 encoders are equalized better than is done in Ref.~\cite{nokkala2021onlineprocessing}.

\subsection{Experimental demonstration of quantum channel equalization}
\label{sec:QCE:borealis}

\begin{figure}[!htb]
	\centering
	\includegraphics[width=\linewidth]{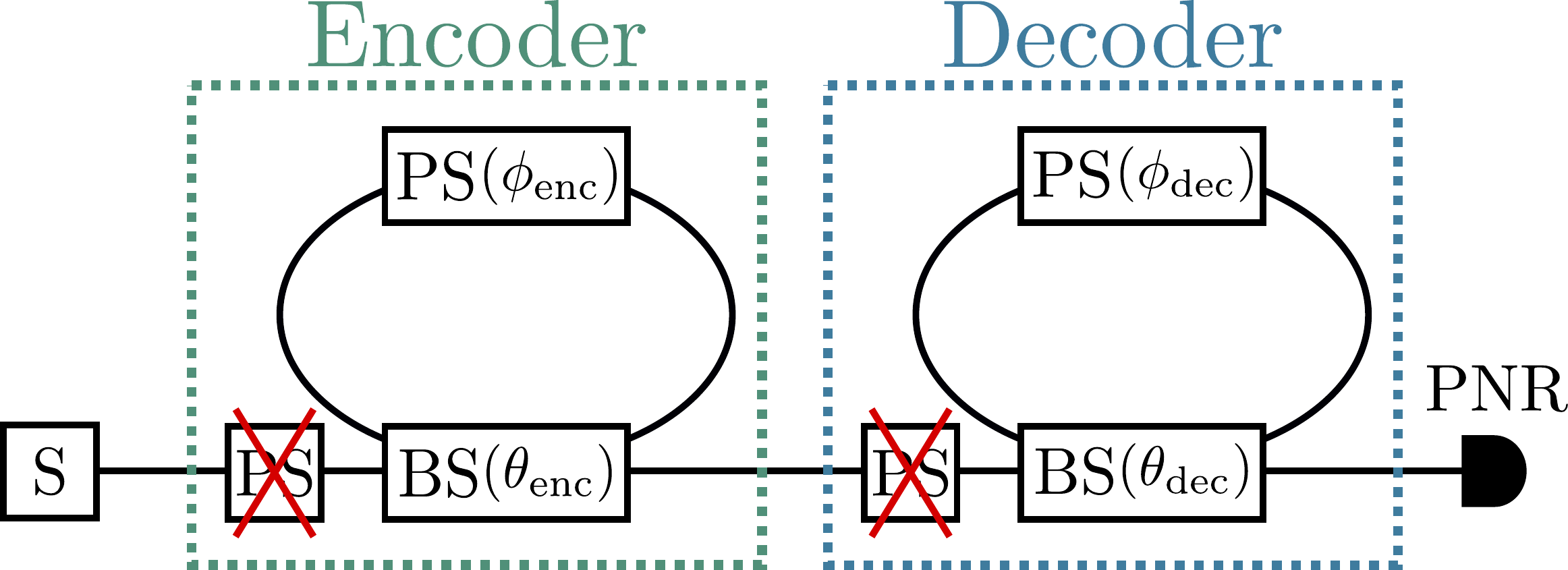}
	\caption{Setup for the QCE task where both the encoder and decoder have a single memory mode. The matrices $\Oenc$ and $\Odec$ are decomposed using beam splitters (BS) and phase shifters (PS). A squeezer (S) is used to generate input states, while the output states are measured using a photon-number-resolving (PNR) detector. As the PSs outside of the loops do not influence the PNR results, they can be disregarded.}
	\label{fig: QCE circuit 1 fb mode}
\end{figure}

To show that our model can be implemented using currently existing hardware, we perform the QCE task on the recently introduced quantum processor Borealis \cite{Borealis}.
Because of hardware restrictions, we only consider the case where $\mfbenc = \mfbdec = 1$. The setup for this task is visualized in \cref{fig: QCE circuit 1 fb mode}.
The input time series consists of squeezed vacuum states (whereas squeezed \textit{thermal} states were used in \cref{sec:QCE:redundancy}).
Both the encoder and decoder are decomposed using beam splitters and phase shifters. Here we use the following definitions for those respective components:
\begin{gather}
    BS(\theta) = e^{\theta (a_i a_j^\dagger - a_i^\dagger a_j)} \\
    PS(\phi) = e^{i \phi a_i^\dagger a_i}
\end{gather}
where $a_i$ ($a_i^\dagger$) is the creation (annihilation) operator on mode $i$. Note that the transmission amplitude of the beamsplitter is $\cos(\theta)$.

Whereas \cref{sec:QCE:redundancy} presented the results of training the decoder, here we will visualize a larger part of the cost function landscape (including sub-optimal decoder configurations).
Note that while evaluating a certain point of the cost function landscape, i.e.~while processing a single time series, the parameters of the beam splitters and phase shifters are kept fixed.
Hence, in \cref{fig: QCE circuit 1 fb mode}, the PNR results are not influenced by the phase shifters outside of the loops (i.e.~outside of the memory modes). These components can be disregarded. Consequently, we can parameterize the encoder (decoder) using only a beam splitter angle $\thetaenc$ ($\thetadec$) and a phase shift angle $\phienc$ ($\phidec$). 

We detect output states with a photon-number-resolving (PNR) detector. In contrast to \cref{sec:QCE:redundancy}, we will not use fidelity as a quality measure, but we will estimate the performance of the QORNN using the following cost function:
\begin{equation}
    \text{cost} = \sum_{k=0}^{K} |\bar{n}_\text{out}^k-\bar{n}_\text{target}^k|~, \label{cost_meanPhot}
\end{equation}
where $\bar{n}_\text{out}^k$ and $\bar{n}_\text{target}^k$ are the average photon numbers of the \textit{actual} output state and the \textit{target} output state at iteration $k$ respectively. $K$ is the total number of states in the input time series.

\begin{figure}[!htb]
	\centering
	\includegraphics[width=\linewidth]{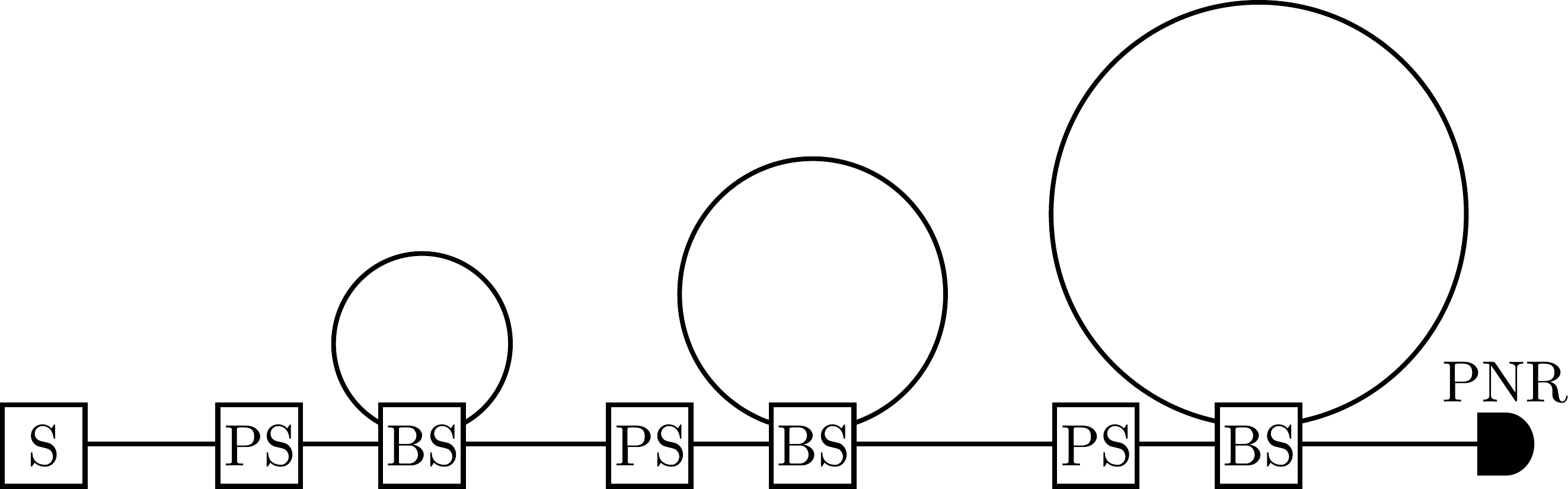}
	\caption{Borealis setup from Ref.~\cite{Borealis}. A squeezer (S) periodically produces single-mode squeezed states, resulting in a train of 216 states that are separated in time by $T = 36\mu s$. These states are sent through programmable phase shifters (PS) and beam splitters (BS). Each BS is connected to a delay line. From left to right in the figure, the lengths of these delay lines are $T$, $6 T$, and $36 T$. The output states are measured by a photon-number-resolving (PNR) detector.}
	\label{fig: borealis circuit}
\end{figure}

The small-scale version of the QCE task (depicted in \cref{fig: QCE circuit 1 fb mode}) is run on the photonic processor Borealis \cite{Borealis}.
The Borealis setup is depicted in \cref{fig: borealis circuit}. 
It consists of a single optical squeezer that generates squeezed vacuum states. These states are sent to a sequence of three dynamically programmable loop-based interferometers of which the delay lines have different lengths, corresponding with propagation times of $T$, $6 T$, and $36 T$ ($T = 36 \mu 
s$). For our experiment, we only use the two leftmost loop-based interferometers. 
More formally, we choose $\theta=0$ for the rightmost BS and $\phi=0$ for the rightmost PS. 

As is explained in more detail in 
\hyperref[sec:methods:QCE:borealis]{Section E of Methods},
we can virtually make the lengths of Borealis' delay lines equal. We do so by lowering the frequency at which we send input states and by putting the $\theta=0$ for the leftmost beam splitter in between input times. 

\begin{figure}[!htb]
\centering
\begin{subfigure}{1\linewidth}
\caption{}
  \includegraphics[width=\linewidth]{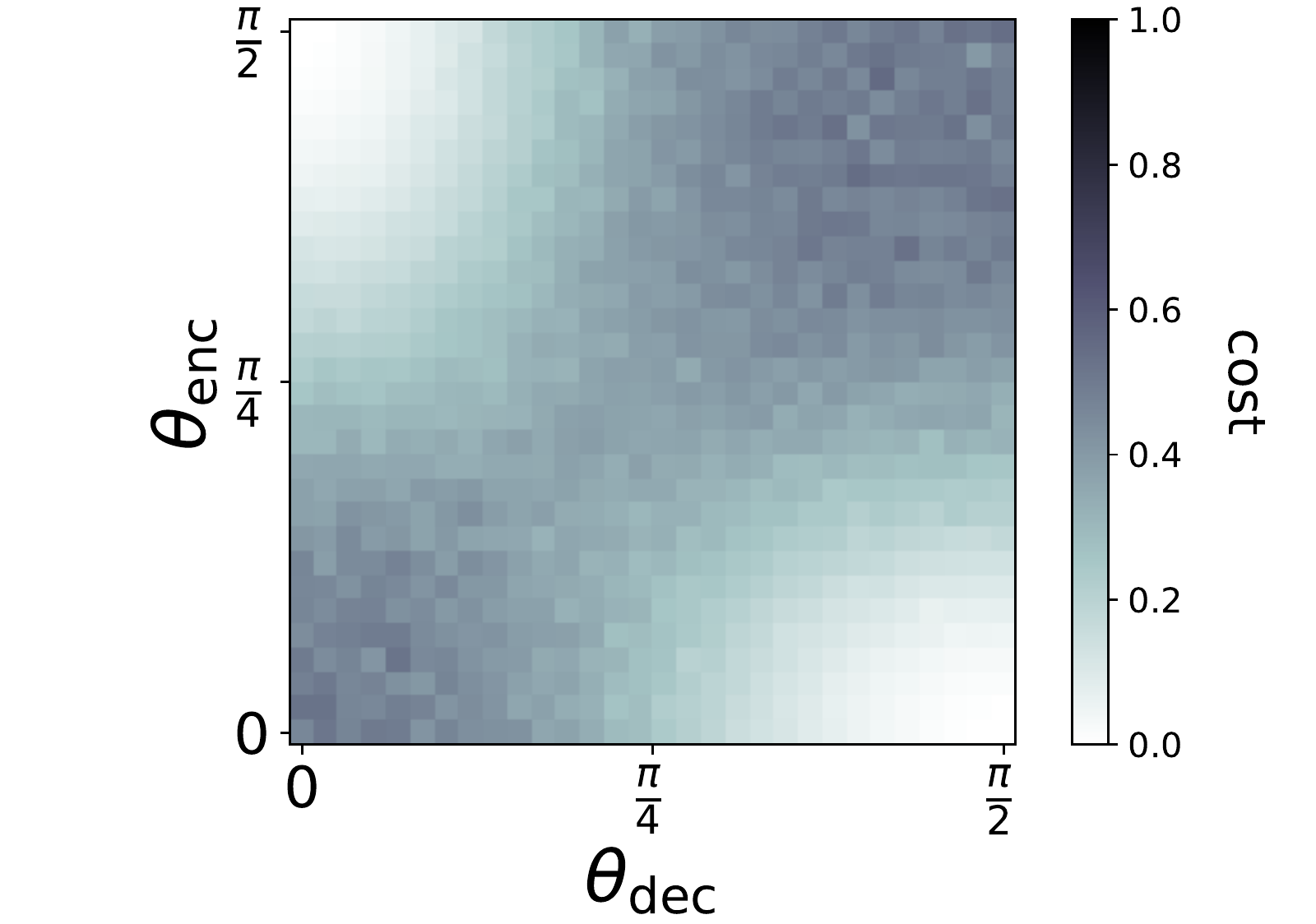}
  \label{fig:borealisSweepBS0BS1:sim}
\end{subfigure}
\begin{subfigure}{1\linewidth}
\caption{}
  \includegraphics[width=\linewidth]{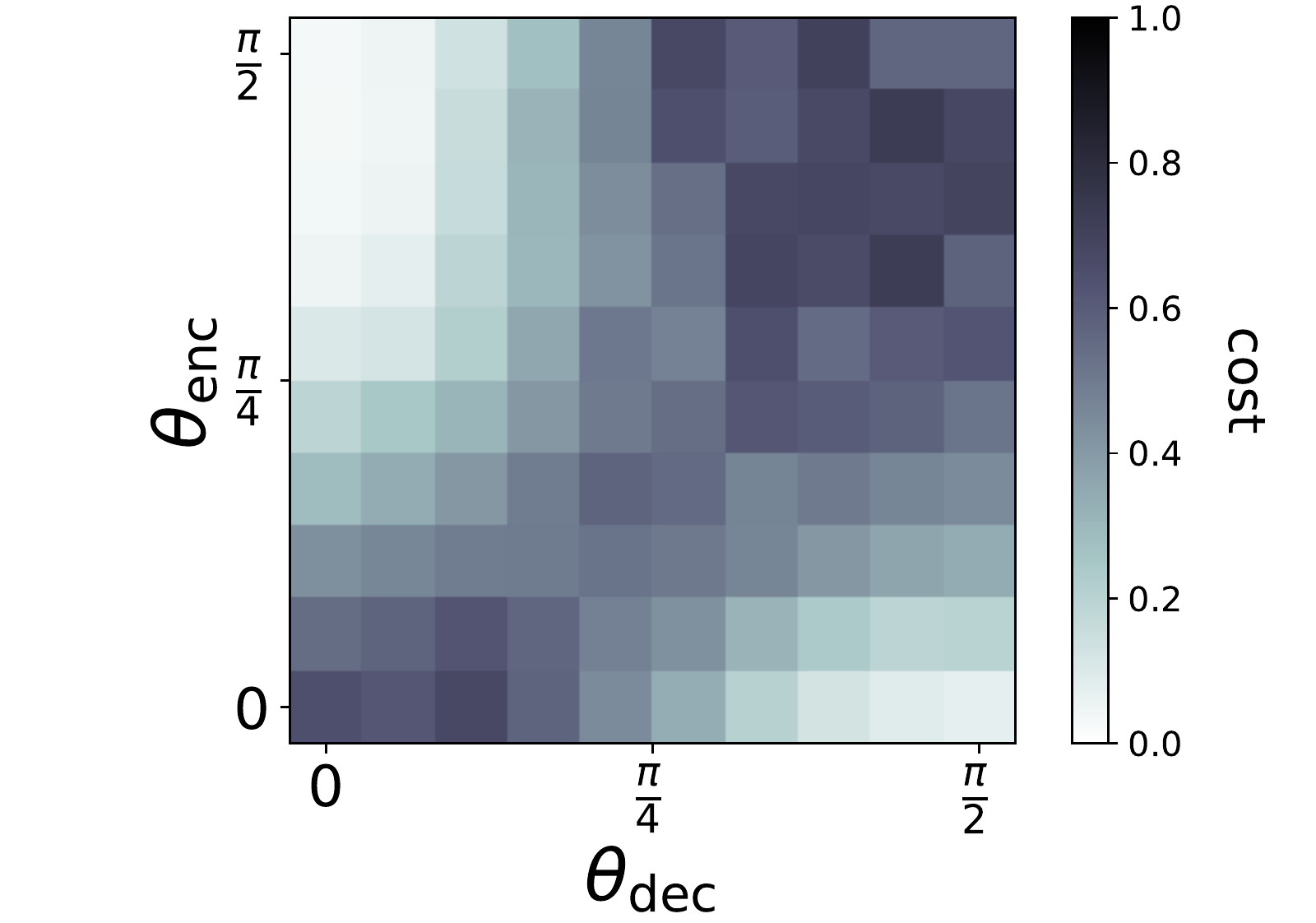}
  \label{fig:borealisSweepBS0BS1:exp}
\end{subfigure}
\caption{
QCE performance of Borealis when $\D=1$ and $\phienc = \phidec = 0$. Fig.~(a) shows the results from numerical simulations, while Fig.~(b) shows experimental results. The cost function is defined in \cref{cost_meanPhot}.
The optimal performance occurs when the input states are delayed by $\D=1$ in the encoder and fully transmitted in the decoder ($\thetaenc=\pi/2$, $\thetadec=0$), or when they are fully transmitted in the encoder and delayed by $\D=1$ in the decoder ($\thetaenc=0$, $\thetadec=\pi/2$).
}
\label{fig:borealisSweepBS0BS1}
\end{figure}

We first consider the case where $\phienc = \phidec = 0$, such that all phase shifters can be disregarded. \cref{fig:borealisSweepBS0BS1} compares the experimental and numerical performance of the QCE task for $\D=1$. We observe that the task is solved perfectly when either the encoder or the decoder delays states by $\D=1$ and the other QORNN transmits states without delay. The performance is worst when both the encoder and decoder delay states with an equal number of iterations (either $\D=0$ or $\D=1$). Indeed, the cost of \cref{cost_meanPhot} is then evaluated between randomly squeezed vacuum states. For beam splitter angles between $0$ and $\pi/2$, we find that the cost follows a hyperbolic surface. We observe good agreement between simulation and experiment.

\begin{figure*}[!htb]
\centering
\begin{tabular}{@{} r M{0.3\linewidth} M{0.3\linewidth} M{0.3\linewidth} @{}}
& \thead{$\theta_{\mathrm{enc}}=0.25\pi$\\$\phi_{\mathrm{enc}}=1.33\pi$}
& \thead{$\theta_{\mathrm{enc}}=0.33\pi$\\$\phi_{\mathrm{enc}}=0.22\pi$} 
& \thead{$\theta_{\mathrm{enc}}=0.50\pi$\\$\phi_{\mathrm{enc}}=0.00$}\\
\begin{subfigure}{0.05\linewidth} \caption{}\label{subfig:a} \end{subfigure} 
  & \includegraphics[width=\linewidth]{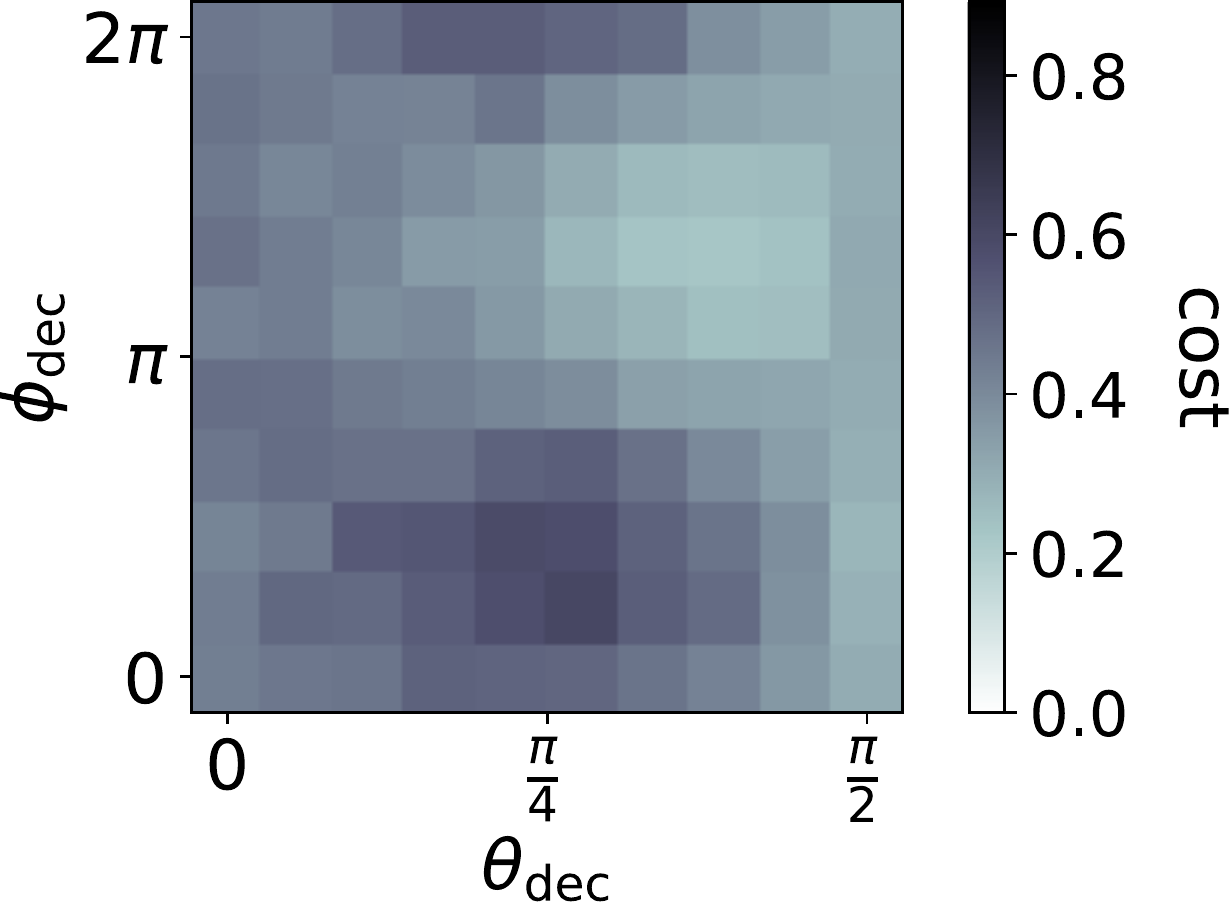}
  & \includegraphics[width=\linewidth]{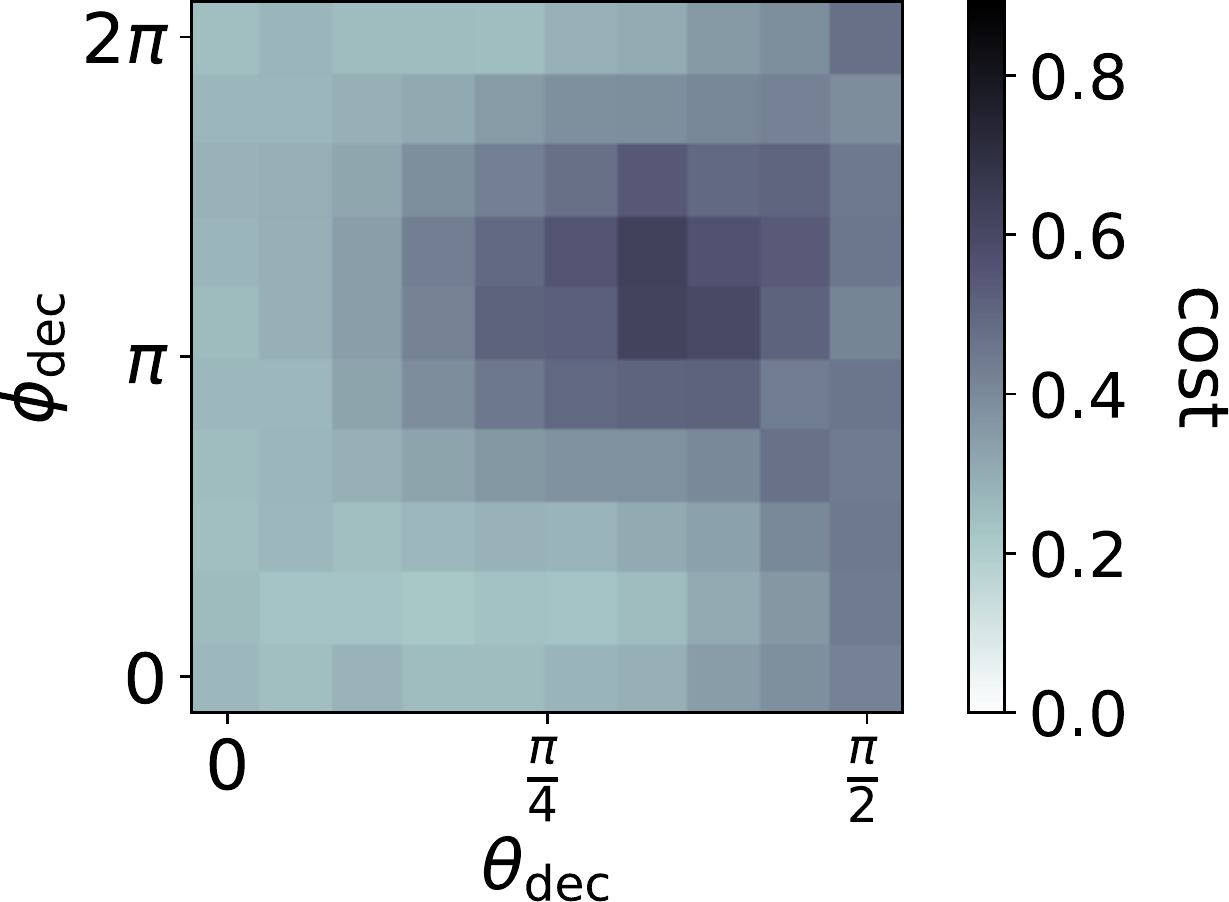}
  & \includegraphics[width=\linewidth]{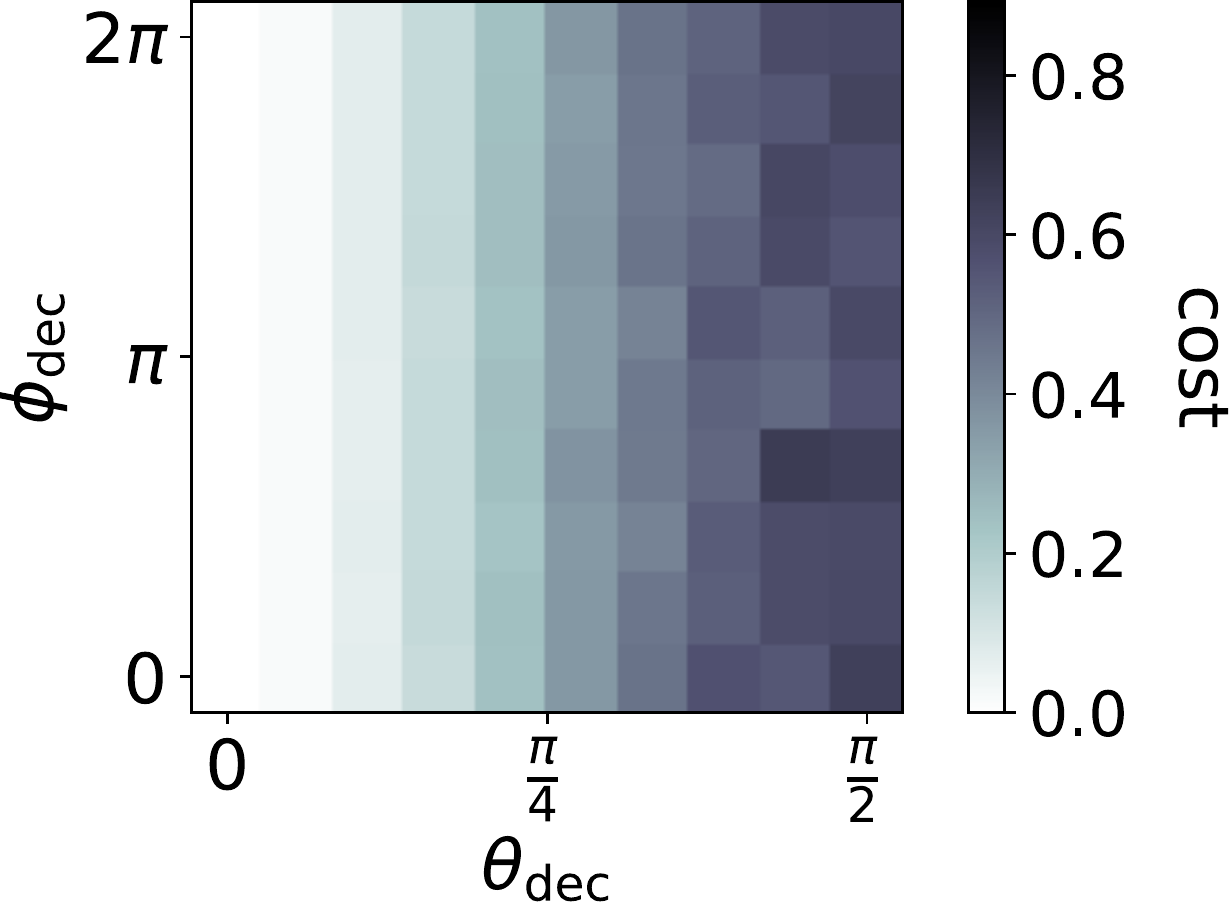}\\ \addlinespace
\begin{subfigure}{0.05\linewidth} \caption{}\label{subfig:b} \end{subfigure} 
  & \includegraphics[width=\linewidth]{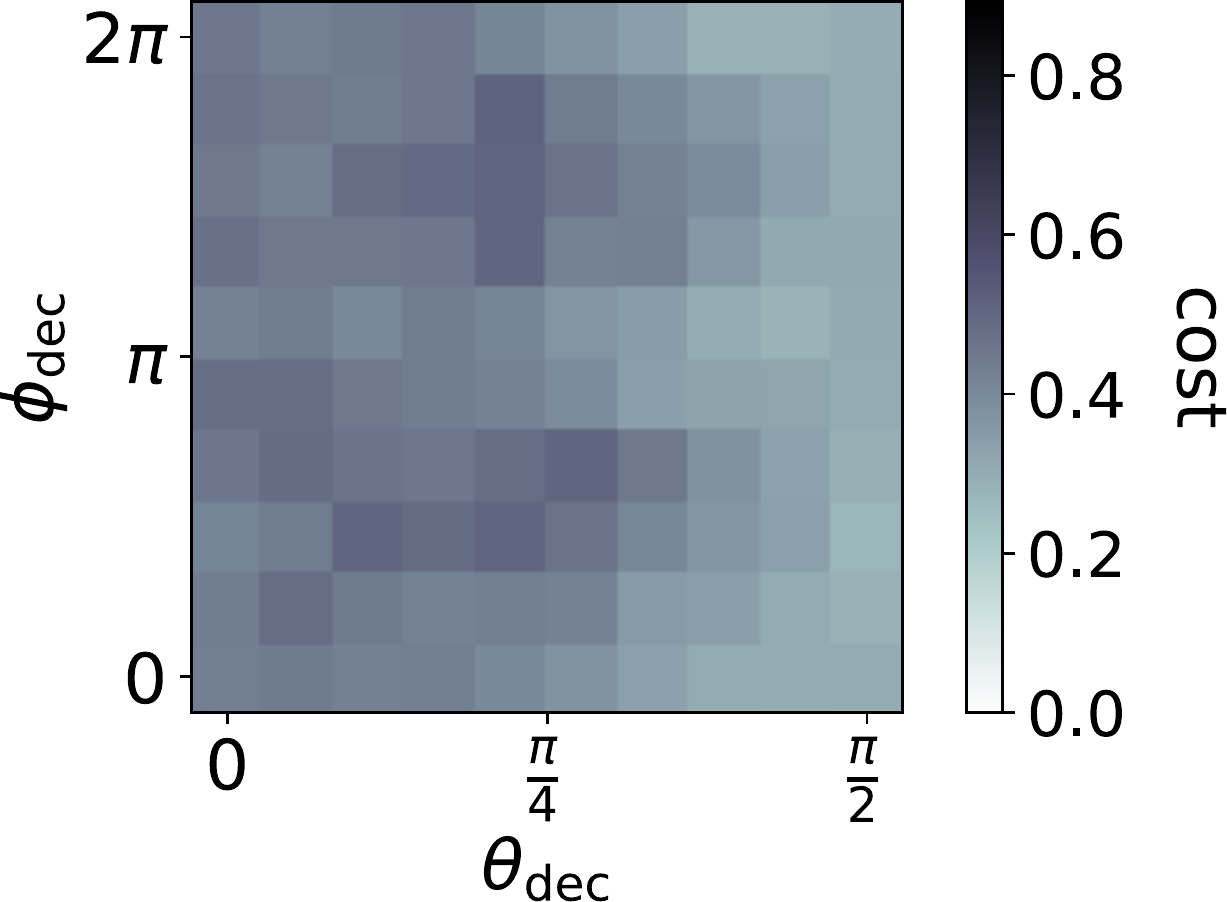}
  & \includegraphics[width=\linewidth]{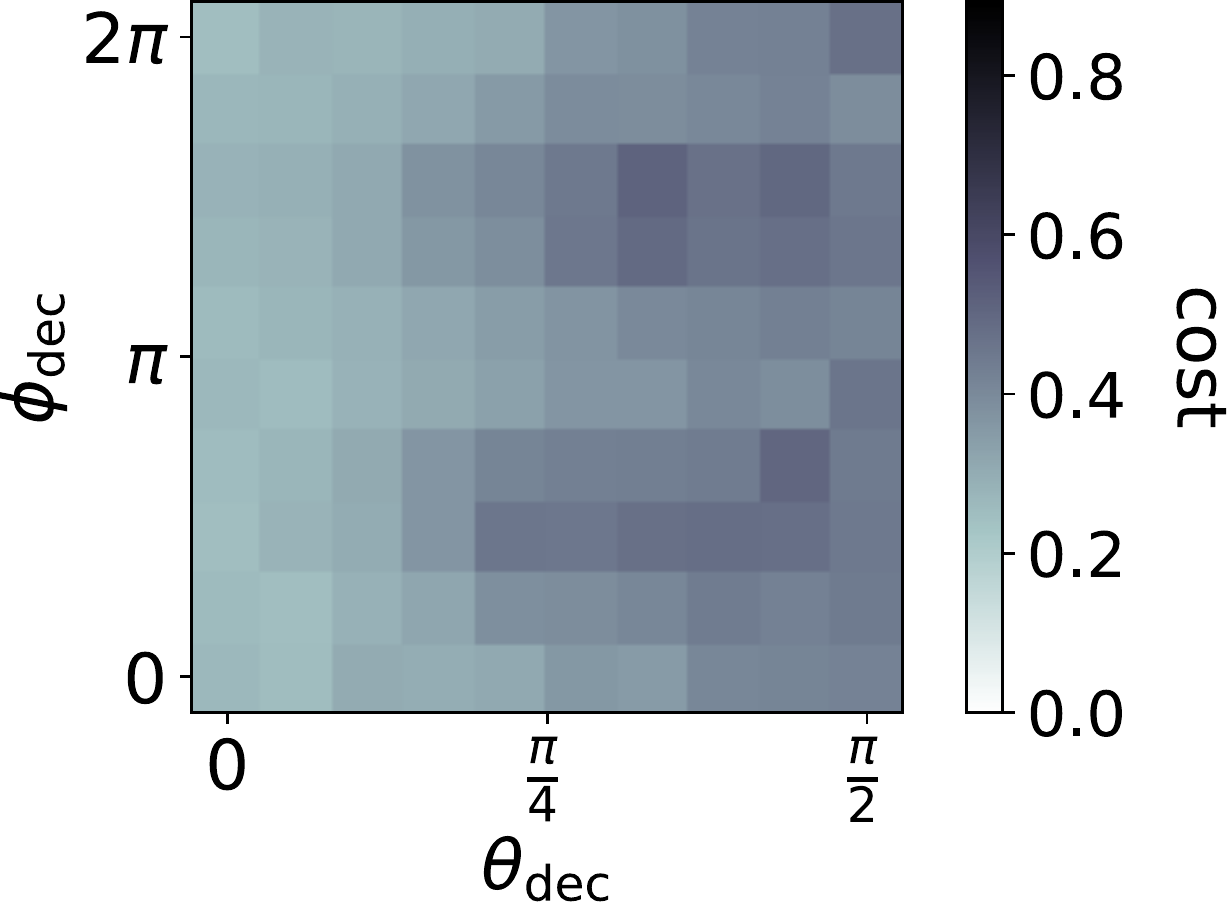}
  & \includegraphics[width=\linewidth]{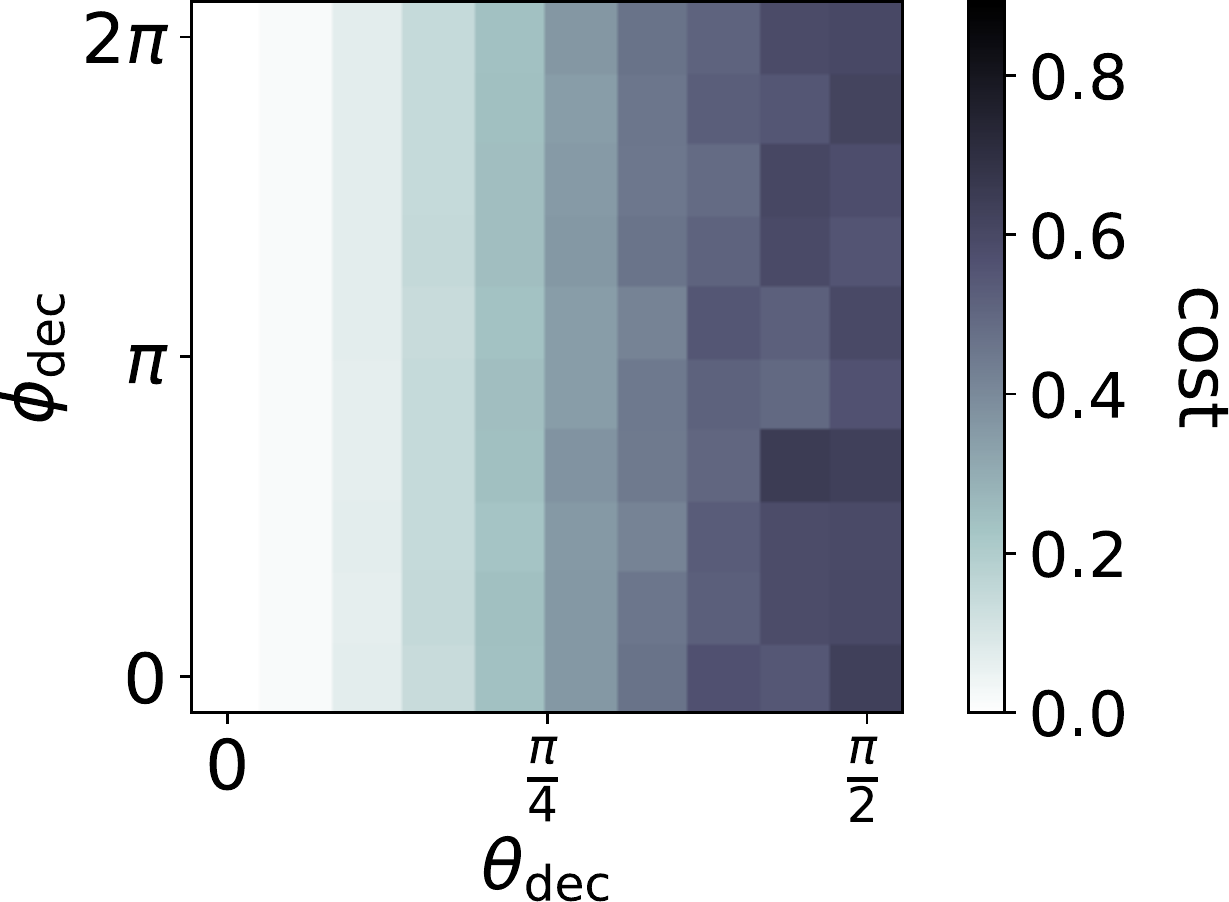}\\ \addlinespace
\begin{subfigure}{0.05\linewidth} \caption{}\label{subfig:c} \end{subfigure} 
  & \includegraphics[width=0.95\linewidth]{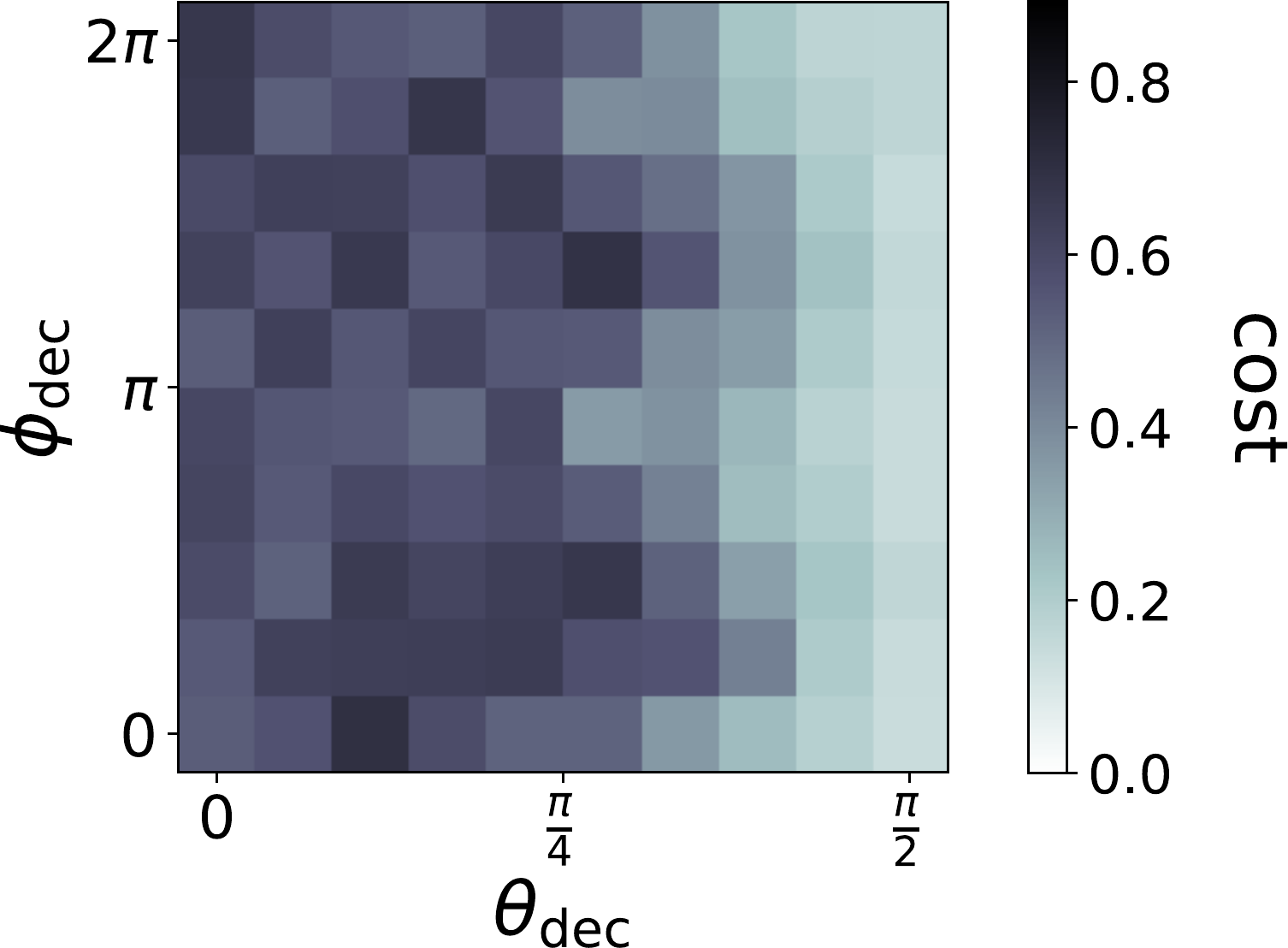} 
  & \includegraphics[width=0.95\linewidth]{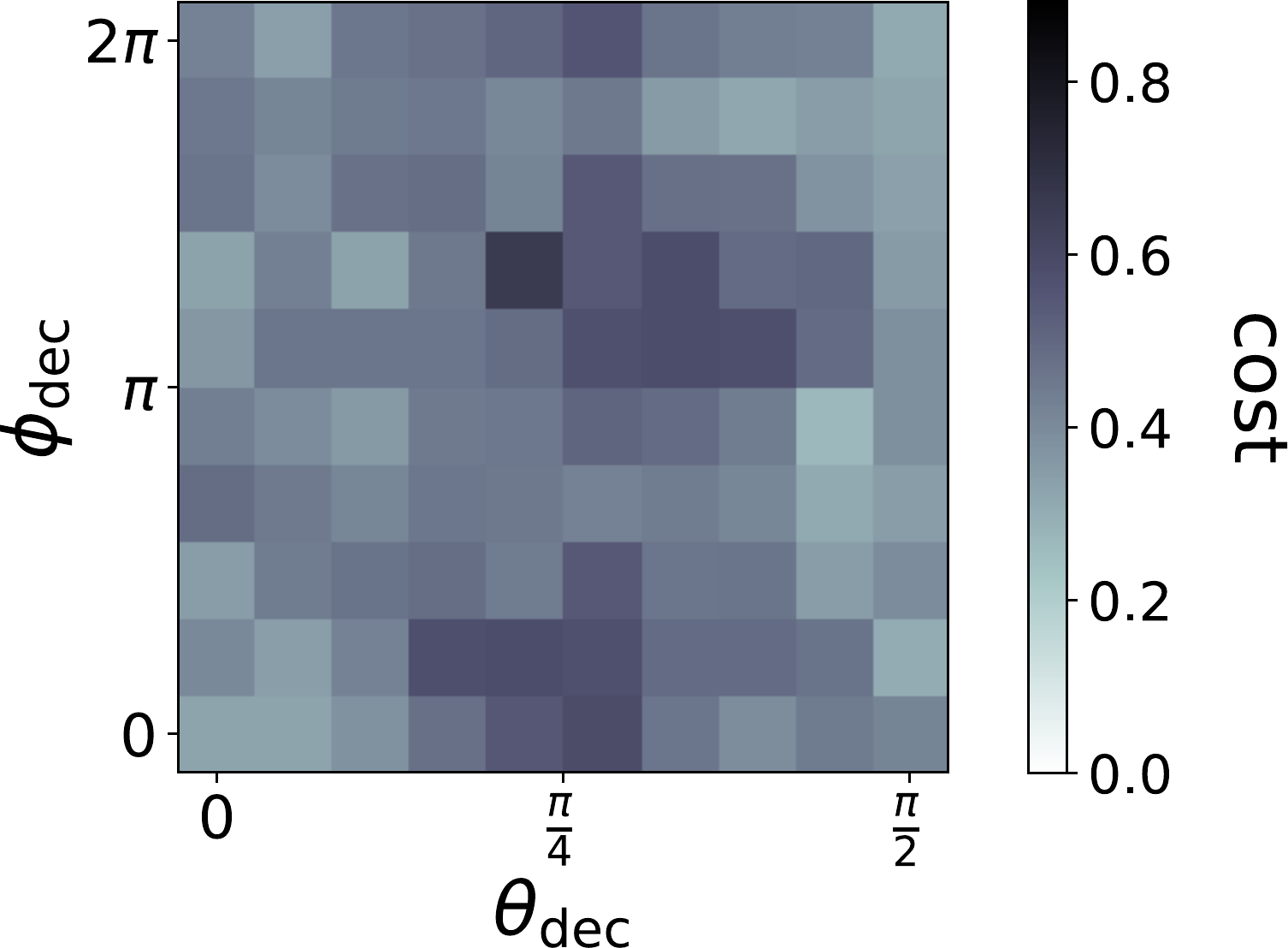}
  & \includegraphics[width=0.95\linewidth]{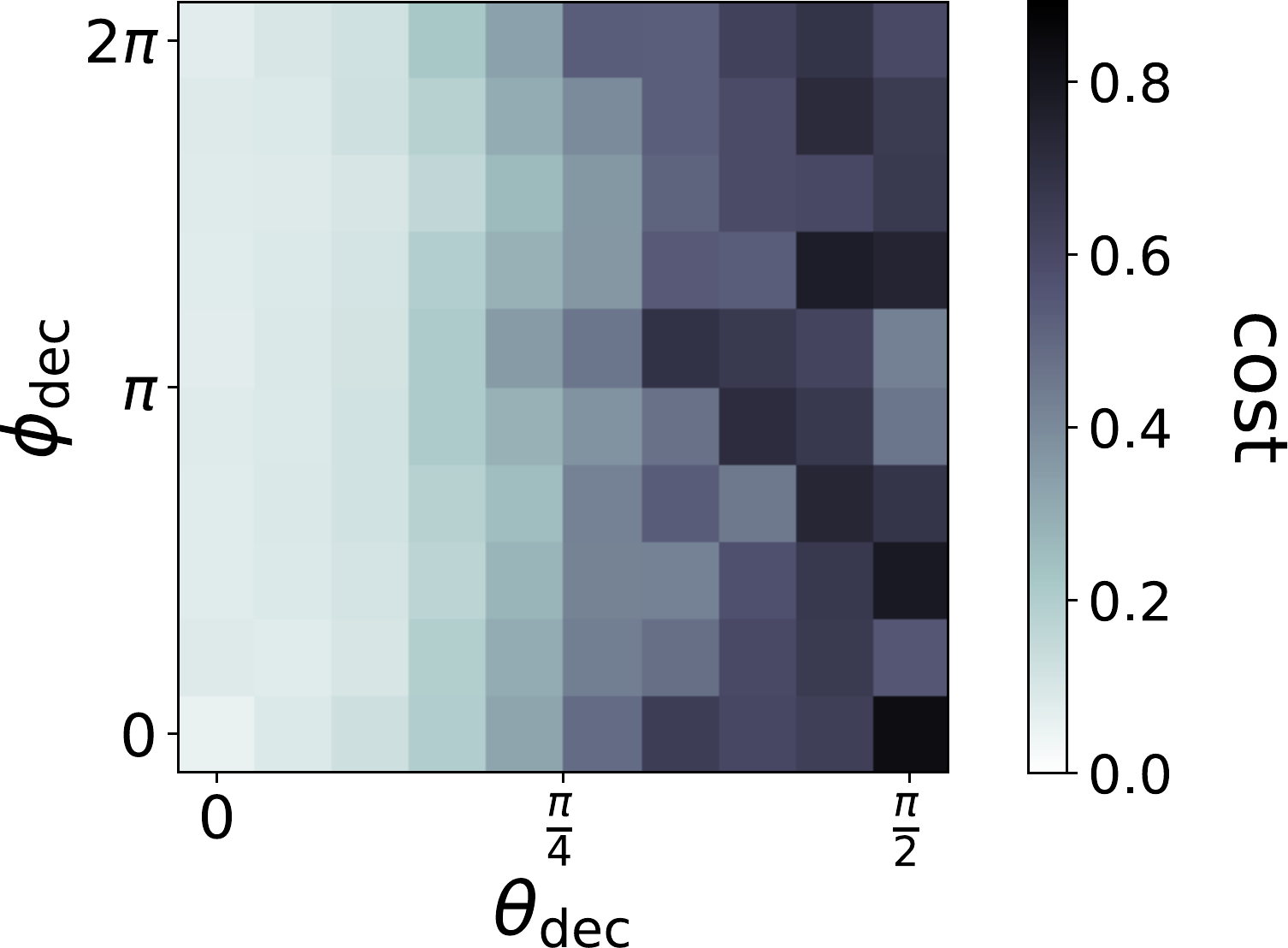}\\ \addlinespace
\end{tabular}
\caption{QCE performance of Borealis when $\D=1$. 
(a) Simulation where phase shifters are tunable without range restriction. (b) Simulation where phase shifters are tunable within $[0,\pi)$.  (c) Experiment where phase shifters are tunable within $[0,\pi)$. The columns correspond with different encoders. Each plot shows the QCE cost (as defined in \cref{cost_meanPhot}) as a function of the decoder parameters $\theta_{\mathrm{dec}}$ and  $\phi_{\mathrm{dec}}$.
}
\label{fig:borealis_results}
\end{figure*}

We now consider the case where $\phienc$ and $\phidec$ are not restricted to 0. Note that the Borealis setup (\cref{fig: borealis circuit}) does not have phase shifters inside the loops. However, as is explained in more detail in \hyperref[sec:methods:QCE:borealis]{Section E of Methods}, we can \textit{virtually} apply the phase shifts $\phienc$ and $\phidec$ \textit{inside} Borealis' loops by dynamically adjusting the phase shifters positioned \textit{before} those loops over time. 
Unfortunately, this procedure is hampered by hardware restrictions. The range of Borealis' phase shifters is restricted to $[-\pi/2,\pi/2]$. This is problematic, since in order to apply a single virtual phase shift value within a specific loop, the phase shifters preceding that loop need to be tuned dynamically across the complete $[-\pi,\pi]$ range. As a result, about half of the parameter values of the phase shifters cannot be applied correctly. When such a value falls outside of the allowed range, an additional phase shift of $\pi$ is added artificially to ensure proper hardware operation.

\cref{fig:borealis_results} shows the QCE ($\D=1$) performance for three different encoders (corresponding to the three columns). We compare classical simulation results (rows 1 and 2) with experimental results (row 3). Whereas the restricted range of Borealis' phase shifters is taken into account in rows 2 and 3, it is not in row 1.
While \cref{fig:borealisSweepBS0BS1} (for $\phienc = \phidec = 0$) showed that the decoder can easily be optimized by considering only $\thetadec = 0$ and $\thetadec = \pi/2$, this optimization is less trivial when $\phienc \neq 0$ and $\phidec \neq 0$. 
We observe that the general trends of the cost function landscapes agree between rows 2 and 3, although row 3 is affected by experimental imperfections.

\section{Conclusions}
\label{sec: conclusions}

We have introduced a new model to process time series of quantum states in real time. We have probed two of its key functionalities: the linear memory capacity and the capability to entangle distinct states within a time series. By comparing with an existing RC-inspired model, we showed that these functionalities benefit from an RNN structure where all internal interactions can be trained. Moreover, the QORNN showed the ability to tackle two \textit{quantum communication} tasks, a domain where optics naturally is the leading hardware platform. First, we demonstrated that the QORNN can enhance the transmission rate of a quantum memory channel with Gauss-Markov noise by providing it with entangled input states. 
The enhancement showed to be near-optimal as compared to previous analytical studies, while the generation scheme of the input states is simpler and can more easily adapt to hardware imperfections than previous approaches. 
Second, we showed that the QORNN can mitigate undesired memory effects in quantum channels. A small-scale experiment of this last task demonstrated the readiness of quantum optics to implement models like the QORNN.

\section*{Methods}
\label{sec:methods}

The classical simulations of our QORNN were carried out using the open-source photonic optimization library \texttt{MrMustard} \cite{MrMustard_github}. This allows us to perform gradient-based optimizations of symplectic circuits and orthogonal symplectic circuits. 

\subsection{STQM}
\label{sec:methods:STQM}
As is explained in Ref.~\cite{nokkala2021onlineprocessing}, a special purpose cost function $f$ can be used to solve the STQM task. $f$ can be defined as a function of the submatrices of the orthogonal symplectic matrix that is associated with the circuit $\mathbf{O}$. With slight abuse of notation, we write the orthogonal symplectic \textit{matrix} that is associated with the \textit{circuit} $\mathbf{O}$ as:

\begin{equation}
\mathbf{O} = 
\left(\begin{array}{ll}
\mathbf{A} & \mathbf{B} \\
\mathbf{C} & \mathbf{D}
\end{array}\right)~.
\end{equation}


In contrast to \cref{eq: env noise cov matrix}, the quadratures are ordered here as follows: $(q_1,p_1,q_2,p_2,q_3,p_3,...)$, where $q_i$ and $p_i$ are the quadratures of mode $i$.

When the delay is nonzero ($\D>0$), Ref.~\cite{nokkala2021onlineprocessing} shows that the goal of the STQM task can be redefined as the following system of equations:
\begin{equation}
\begin{cases}
\mathbf{D} \approx \mathbf{0}~, \\
\mathbf{C A}^{\D-1} \mathbf{B} \approx \mathbf{I}~, \\
\mathbf{C A}^{t} \mathbf{B} \approx \mathbf{0}~,~ \forall \, t \neq \D-1~.
\end{cases}
\end{equation}

Hence, we choose the following cost function:
\begin{equation}
\begin{aligned}[b]
f\left(\mathbf{U}\right) =& \; 0.5\|\mathbf{D}\|+5\left\|\mathbf{C A}^{\D-1} \mathbf{B}-\mathbf{I}\right\| \\ +& \; 0.5\sum_{\substack{0 \leq t < T \\ t \neq \D-1}}{\|\mathbf{C}\mathbf{A}^t \mathbf{B}\|},
\end{aligned}
\label{eq:special_purpose}
\end{equation}

where $\|\cdot\|$ is the Frobenius norm. Note that the last sum in \cref{eq:special_purpose} was not used in Ref.~\cite{nokkala2021onlineprocessing}, as these terms appeared to worsen the results. However, we have found that their inclusion is beneficial in our work. A numerical parameter sweep showed that the factors 0.5, 5, and 0.5 for the terms in \cref{eq:special_purpose}, together with a value of $T=10$ yield good results.

The learning rate is initialized at a value of 0.01 and decreased throughout the training procedure until the cost function converges. 
For each value of $(\mout,\mfb)$ in \cref{fig: STQM sweep}, the training is repeated for an ensemble of 100 initial conditions of the QORNN.
After training, a test score is calculated as the average fidelity over a test set of 100 states. Only the lowest test score in the ensemble of different initial conditions is kept, as this corresponds to a degree of freedom that is exploitable in practice.

In order to evaluate the test score, squeezed thermal states are used as input. 
In the case of a single mode, we first generate a thermal state with an expected number of photons $\bar{n}_\text{th}$. 
Afterwards, the state is squeezed according to a squeezing magnitude $r_\text{sq}$ and a squeezing phase $\phi_\text{sq}$.
The relevant parameters of this state generation process are chosen uniformly within the following intervals: $\bar{n}_\text{th} \in [0,10)$, $r_\text{sq} \in [0,1)$ and $\phi_\text{sq} \in [0,2\pi)$.

If $\mout>1$, the state generation is altered as follows. We first generate a product state of single-mode thermal states. The $M$-mode product state is then transformed by a random orthogonal symplectic circuit, followed by single-mode squeezers on all modes. A second random orthogonal symplectic circuit transforms the multi-mode state to the final input state.

\subsection{Entangler task}
We evaluate the cost function as follows.
We send vacuum states to the QORNN for 10 iterations, such that the model undergoes a transient process. We numerically check that this transient process has ended by monitoring the convergence of the average photon number in the output mode. We then calculate the logarithmic negativity of the 2-mode state formed by the output at iteration $10$ and $10+S$.

The logarithmic negativity of a 2-mode state is calculated from the determinants of the state's covariance matrix (and its submatrices) as described in Ref.~\cite{olivares2012quantum}. Note that we do not calculate the negativity from the symplectic eigenvalues of the covariance matrix (as is done for example in Ref.~\cite{vidal2002computable}), which is beneficial for numerical optimization.

The cost function is defined as the negative of the logarithmic negativity. The training is performed using 4000 updates of $\Sb$ and with a learning rate of 0.01. For each value of $(S,\mfb)$ in \cref{fig:ENT_results}, the training is repeated for an ensemble of 100 initial conditions of $\Sb$. Again, only the lowest score in the ensemble is kept.

\subsection{Superadditivity task}

For more details on the calculation of the optimal transmission rate of the quantum channel with Gauss-Markov noise, both with and without entangled input states, we refer to Ref.~\cite{schafer2011GaussMarkov2_TypeB}.

For the case of entangled input, the optimization problem defined by Eq. 9 of this Reference is solved under the restriction that the covariance matrix $\bm{\gamma}_\text{in}$ is produced by the QORNN. $\bm{\gamma}_\text{mod}$ is a general covariance matrix that takes into account the required input energy constraint (Eq. 13 in Ref.~\cite{schafer2011GaussMarkov2_TypeB}). 

The cost function is defined as the negative of the transmission rate. The training is performed using 1000 updates of $\Sb$ and with a learning rate of 0.5.

\subsection{Quantum channel equalization}
\label{sec:methods:QCE:redundant}
In contrast to the STQM task, here we use the negative of the fidelity (between the desired output and the actual output) as a cost function, both during training and testing. 
The input states are single-mode squeezed thermal states where $\bar{n}_\text{th}$, $r_\text{sq}$, and $\phi_\text{sq}$ are chosen uniformly within the following ranges: $\bar{n}_\text{th} \in [0,10)$, $r_\text{sq} \in [0,1)$ and $\phi_\text{sq} \in [0,2\pi)$.

$\Oenc$ and $\Odec$ are initialized randomly.
Every epoch, 30 squeezed thermal states are sent through the QORNNs. 
In order to account for the transient process at the start of each epoch, the fidelity is only averaged over the last 20 output states.

The training is performed using 2000 updates of $\Odec$ and with a learning rate of 0.05. In \cref{fig: QCE fid vs Ienc}, the training is repeated for an ensemble of 20 initial conditions of $\Sb$  for every value of $(\D,\mfbdec)$ and for each encoder initialization.
After training, a test score is calculated by simulating a transient process during 10 iterations and averaging the fidelity over 100 subsequent iterations.
The lowest test score in each ensemble is kept.

\subsection{Experimental demonstration of quantum channel equalization}
\label{sec:methods:QCE:borealis}

This section describes the technical details of both the simulations and the experiments shown in \cref{sec:QCE:borealis}. We first describe how the QORNN can be mapped onto the Borealis setup. Afterwards, we describe the input state generation, the post-processing of the PNR results, and some other relevant parameters for the simulations and experiments.

\subsubsection{Mapping the QORNN model on Borealis}
We map our model on the Borealis setup of \cref{fig: borealis circuit}. We choose $\phi=0$ and $\theta=0$ for the rightmost phase shifter and beam splitter respectively, such that these components and their corresponding delay line can be disregarded.
After applying these changes, it is clear that the Borealis setup differs from the setup in \cref{fig: QCE circuit 1 fb mode} in two ways. First, the remaining delay lines of the Borealis setup have different lengths, which is not the case in \cref{fig: QCE circuit 1 fb mode}. Second, Borealis does not have phase modulators in the delay lines. We can circumvent both of these problems as follows.

First, we can virtually increase the length of the leftmost delay line (i.e.~the encoder delay line) from $T$ to $6T$ by performing two changes: (1) we lower the frequency at which we input states from $\frac{1}{T}$ to $\frac{1}{6T}$ and (2) we store the encoder memory state as long as we do not input a new state.
More formally, we do the following.
Before generating a new input state, we put the values of the beam splitter angles to $\theta_\text{enc}$ (for the encoder) and $\theta_\text{dec}$ (for the decoder). Once a state enters the encoder, it interacts with the memory state, leading to an output state and a new memory state for the encoder. The output state of the encoder is sent to the decoder. 
We now put $\thetaenc=0$ for a period of $6T$, such that no new inputs can enter the loop.
Hence, the new memory state is stored in the delay line of the encoder. During this period, no states are generated and no interactions occur between the encoder's memory mode and input/output mode. Afterward, a new state is generated and the process is repeated. 

Second, we can apply virtual phase shifts in the delay lines of \cref{fig: borealis circuit} by using the phase shifters in front of the loops. To do this, we utilize an existing functionality of Borealis' control software. This functionality (implemented in \texttt{StrawberryFields} \cite{StrawberryFields_github} as \texttt{compilers.tdm.Borealis.update\_params}) is normally used to compensate for unwanted phase shifts in the delay lines. 
Given such a static unwanted phase shift, the compensation is performed by adjusting the phase shifters in front of the loops \textit{over time}, such that they apply different phase shifts at different iterations. 
The actual unwanted phase shifts that are present in the hardware can be measured before running an experiment. 
We now choose to operate the setup \textit{as if} there were phase offset sets $\phi_\text{1,unwanted} - \phi_\text{enc}$ and $\phi_\text{2,unwanted} - \phi_\text{dec}$ in the first two delay lines respectively.
Hence, we apply net virtual phase shifts in the loop with values $\phi_\text{enc}$ and $\phi_\text{dec}$. 

Unfortunately, this procedure is undercut by a hardware restriction of Borealis. The range of these phase shifters is limited to $[-\pi/2,\pi/2]$, which means that about half of the desired phase shifts cannot be implemented correctly. When a phase shift falls outside of the allowed range, an additional phase shift of $\pi$ is added artificially to ensure proper hardware operation.
\cref{code:updateparams} shows the Python code for this process.

\begin{listing}[!htb]
\raggedleft
\begin{minted}[frame=lines,fontsize=\footnotesize]{python}
import numpy as np
def update_PSs(PS_params, measured_phis, virtual_phis):
    '''
    PS_params [arr,arr,arr]: parameters of the 3 phase 
    shifters before the loops at all 216 iterations
    (i.e. each array consists of 216 floats)
    measured_phis [float,float,float]: unwanted phase 
    shifts measured inside the loops
    virtual_phis [float,float,float]: net phase shifts 
    we want to apply inside the loops
    '''
    delays = [1, 6, 36] # relative lengths of the loops
    PS_params_updated = []
    corr_previous_loop = np.zeros(216)
    for i, delay in enumerate(delays):
        # calculate new parameter values for loop i
        offset = measured_phis[i] - virtual_phis[i]
        corr_loop = np.arange(216) // delay * offset
        phi_corr = PS_params[i]
        phi_corr += corr_loop - corr_previous_loop

        # map values within [-np.pi/2, np.pi/2]
        phi_corr = np.mod(phi_corr, 2*np.pi)
        phi_corr = np.where(phi_corr > np.pi,\
                            phi_corr - 2*np.pi, phi_corr)
        phi_corr = np.where(phi_corr < -np.pi/2,\
                            phi_corr + np.pi, phi_corr)
        phi_corr = np.where(phi_corr > np.pi/2,\
                            phi_corr - np.pi, phi_corr)

        PS_params_updated.append(phi_corr)
        corr_previous_loop = corr_loop
    return PS_params_updated
\end{minted}
\caption{Code to apply virtual phase shifts inside Borealis' loops. 
This is an adapted version of the function \texttt{compilers.tdm.Borealis.update\_params} of \texttt{StrawberryFields} \cite{StrawberryFields_github}.
}
\label{code:updateparams}
\end{listing}

\subsubsection{Generating input states, post-processing PNR results, and simulation parameters}
\label{sec:methods:borealis:part2}

Both for the computer simulations and for the experiments presented in \cref{sec:QCE:borealis}, the input states are generated by a single-mode optical squeezer that operates on a vacuum state. In every iteration, the squeezing magnitude is chosen randomly between either 0 or 1, effectively encoding bit strings into the quantum time series. 
It is worth noting that public online access to Borealis does not allow for multiple squeezing magnitudes within one experiment. Our experiments were therefore performed on-site. 

The output of Borealis consists of the average photon numbers gathered over 216 iterations. Due to the mapping described in the previous section, only 36 of these average photon numbers can be used. 
To account for the transient process, the results of the first 10 of these 36 output states are disregarded. The PNR results of the remaining 26 output states are used to estimate the decoder's QCE performance.

For the \textit{simulations}, 10 runs (each using a different series of input states) are performed per set of gate parameter values of the QORNNs. The cost is therefore averaged over 260 output states. For the \textit{experiments} shown in \cref{fig:borealisSweepBS0BS1} and \cref{fig:borealis_results}, the number of runs per data point is 7 and 2 respectively.

In hardware experiments, non-idealities occur that are not considered in the classical simulations. We compensate for two such non-idealities by re-scaling the experimentally obtained average photon numbers. On the one hand, there are optical losses. On the other hand, the optical squeezer suffers from hardware imperfections that lead to effective squeezing magnitudes that decrease over time when the pump power is kept constant.
The re-scaling is performed as follows.
At each iteration, we calculate the average of the number of detected photons over all experiments. We fit a polynomial of degree 2 to these averages. Given an experimental result, i.e.~a series of average photon numbers, we then divide each value by the value of the polynomial at that iteration. The resulting series is rescaled such that the average of the entire series matches the average value of the entire series of input states. 

\section*{Acknowledgments}
We would like to thank Filippo M. Miatto for his insights, feedback, and assistance with software-related challenges.
We are also grateful to Johannes Nokkala for sharing his expertise, and to Lars S. Madsen, Fabian Laudenbach, and Jonathan Lavoie for making the hardware experiment possible.
This work was performed in the context of the Flemish FWO project G006020N and the Belgian EOS project G0H1422N. It was also co-funded by the European Union in the Prometheus Horizon Europe project.

\bibliographystyle{IEEEtran}
\bibliography{main}

\end{document}